\documentclass[
 aps,
 physrev,
 reprint,
 amsmath,
 amssymb
]{revtex4-2}

\usepackage{xcolor}

\usepackage{graphicx}
\usepackage{placeins}
\usepackage{dcolumn}
\usepackage{bm}
\usepackage[normalem]{ulem}
\usepackage{cancel}
\usepackage[pagebackref=false,colorlinks,linkcolor=blue,filecolor=green,urlcolor=blue,citecolor=blue]{hyperref}

\usepackage{comment}
\usepackage{amsmath}

\begin{document}

\title{Switching off a high-harmonic polarization channel by anti-unitary symmetry}

\author{Milad Jangjan}
\email{milad.jangjan@uni-rostock.de}

\author{Francisco Navarrete}
\author{Dieter Bauer}

\affiliation{Institute of Physics, University of Rostock, 18051 Rostock, Germany}

\begin{abstract}
We study polarization-resolved high-harmonic generation in a laser-driven two-leg Su--Schrieffer--Heeger ladder with asymmetric dimerization under periodic boundary conditions. A field linearly polarized along the ladder can generate both longitudinal and transverse radiation, whose harmonic content is controlled by lattice symmetry and electronic filling. Using symmetry analysis of the time-dependent density matrix, we show that reflection symmetry enforces odd-order harmonics in the longitudinal response and, when the transverse response is nonzero, even-order harmonics in the transverse direction. At half filling, an additional anti-unitary symmetry relating the occupied and empty bands suppresses the transverse current completely. A unitary chiral symmetry in the clean ladder relates the same bands but fails to enforce this cancellation, showing that band swapping alone is insufficient: the suppression requires both a density-matrix constraint that persists during the driven evolution and the appropriate transformation of the transverse-current operator. Moving away from half filling lifts this cancellation and allows transverse emission, while the reflection selection rule remains whenever reflection symmetry is preserved. Numerical calculations at quarter and half filling, combined with onsite potentials that selectively break the relevant symmetries, confirm these two distinct mechanisms. Thus, reflection symmetry determines which harmonic orders occur, whereas the interplay of anti-unitary symmetry and filling controls whether the transverse HHG channel is present at all.
\end{abstract}
\maketitle

\section{Introduction}

High-harmonic generation (HHG) in solids has been a powerful technique for investigating electron dynamics and the microscopic structure of materials~\cite{Ghimire2011Observation,Hohenleutner2015RealTime,Garg2016Multi}. The field has rapidly expanded since its first observation in bulk ZnO in 2011~\cite{Ghimire2011Observation}. When a strong laser field drives a crystalline solid, the resulting nonlinear electronic motion can emit radiation at integer multiples of the driving frequency~\cite{Ghimire2011Observation,Schubert2014Subcycle,Luu2015Extreme,Ciappina2025SolidState,navarrete2019crystal}. The properties of this emission, such as its cutoff energy, intensity, and polarization, are highly sensitive to the band structure, symmetries, and scattering mechanisms of the material~\cite{Vampa2015Theoretical,Vampa2015Linking,Lorenzo2026HighHarmonic,Luu2018Measurement}.

In gases, HHG is commonly understood in terms of electrons that tunnel from the parent ion, gain energy from the laser field, and recombine while emitting high-frequency radiation. In crystalline solids, the electronic motion is shaped by the band structure: laser-driven electrons move through crystal momentum, can be transferred between filled and empty bands, and generate a time-dependent microscopic current whose Fourier spectrum gives the emitted harmonics. Different theoretical descriptions emphasize different aspects of this process, but in the present work we focus on symmetry constraints on the total current.
For such a process, selection rules determine which harmonic orders appear and are dictated by the dynamical symmetries of the system~\cite{Alon1998SelectionRules,Neufeld2019FloquetGroup,Neufeld2019Symmetries,Liu2016SelectionRules,Jangjan2026NESS,Moiseyev2015SelectionSolids,ceccherini2001dynamical,Ciappina2024HighOrder,Kozak2025Noncollinear,Kraus2024Enhancing,Kuzkova2025Boosting}. Notably, even-order harmonics are generally forbidden in centrosymmetric systems, since inversion symmetry combined with the half-period anti-periodicity of the drive enforces odd-only emission. More generally, dynamical symmetries can constrain not only the harmonic order, but also the direction and polarization of the emitted radiation. This classification of dynamical symmetries and their associated selection rules was developed in Refs.~\cite{Alon1998SelectionRules, Neufeld2019FloquetGroup, Moiseyev2015SelectionSolids}; here we derive the corresponding reflection-based rule explicitly for the two-leg ladder geometry, while identifying a distinct anti-unitary mechanism, tied to electronic filling, that switches an entire polarization channel on or off rather than merely constraining its harmonic content.

A simple yet rich platform for exploring these nonlinear phenomena is the Su--Schrieffer--Heeger (SSH) model, originally developed to describe polyacetylene~\cite{su1979solitons}. While traditionally studied for its dimerized ground state, its simple two-site unit cell and well-understood band structure make it an ideal testbed for investigating laser-driven electron dynamics. Recent studies have shown that the SSH model exhibits characteristic HHG signatures, with harmonic spectra that reflect the underlying lattice geometry~\cite{Bauer2018High,jurss2019high,zhang2024high}.
Bulk tight-binding formulations of HHG in SSH-type and related two-band models have also clarified the importance of Brillouin-zone integration, topology, and current operators consistent with the chosen Bloch representation~\cite{drueke2019robustness,moos_intense-laser-driven_2020}.
Beyond the standard SSH chain, numerous generalized versions have been explored in the context of HHG. These include extended SSH models with longer-range hoppings that host multiple topological phases~\cite{bera2023topological,Ma2022PRB106125117}, SSH-Hubbard models incorporating electron-electron interactions~\cite{pooyan2024harmonic,marques2016multihole}, Holstein-SSH models with electron-phonon coupling leading to polaron formation~\cite{caceres2025polaronic}, and disordered SSH chains displaying multiple-plateau structures arising from spectral interference~\cite{cui2024multiplateau}.

Extending the SSH model to more complex geometries, such as ladder configurations, introduces additional degrees of freedom and richer band structure. In a two-leg ladder SSH model, inter-chain hopping terms couple two SSH chains, resulting in a system that can host multiple topological phases depending on the interplay between intra-chain dimerization and inter-chain coupling~\cite{jangjan2020floquet}.
Here we use this ladder geometry for a different purpose: it provides a minimal four-band bulk system in which the longitudinal and transverse HHG currents are constrained differently by symmetry.
In addition to lattice symmetries dictating the allowed harmonics in solid-state HHG through dynamical selection rules~\cite{Liu2016SelectionRules,Moiseyev2015SelectionSolids}, the electronic filling level or doping of a system has been shown to qualitatively affect HHG spectra. For example, theoretical studies using a Hubbard chain demonstrate that moving away from half filling modifies the HHG yield and spectral features, highlighting that filling is a critical parameter influencing harmonic emission in driven lattice systems~\cite{Hansen2022Doping}.
The filling effect considered below is more specific. At half filling of the clean ladder, the two initially filled bands and the two initially empty bands are mapped into each other by a combined reflection--particle-hole anti-unitary symmetry. This gives an additional constraint on the time-dependent density matrix and can cancel a whole current component, rather than merely changing the relative strength of harmonic peaks.
Despite these advances, HHG in extended SSH models, such as two-leg ladder configurations, remains relatively unexplored.

In this work, we study high-harmonic generation in both the
longitudinal and transverse directions of a laser-driven two-leg SSH ladder under periodic boundary conditions (PBCs) in the velocity gauge, focusing exclusively on bulk dynamics. Using symmetry analysis of the time-dependent density matrix, we derive symmetry-based selection rules for the polarization-resolved HHG spectra. Two distinct symmetry consequences emerge. A reflection that reverses the coordinate along the ladder, combined with the half-period sign reversal of the drive, enforces odd-order harmonic emission in the longitudinal direction and, whenever the transverse current is nonzero, even-order emission in the transverse direction. At half filling, an additional symmetry relating the occupied and empty bands forces the transverse current to vanish identically. Away from half filling, this cancellation is lifted, while the reflection-based selection rule remains whenever the corresponding reflection symmetry is preserved.
The remainder of this paper develops these two mechanisms in turn: a reflection-based selection rule on harmonic order, and a distinct, filling-dependent anti-unitary constraint on whether the transverse channel is active at all. These analytical predictions are confirmed by numerical time evolution of the density matrix under periodic boundary conditions.

\begin{figure}[t]
 \centering
\includegraphics[width=\columnwidth]{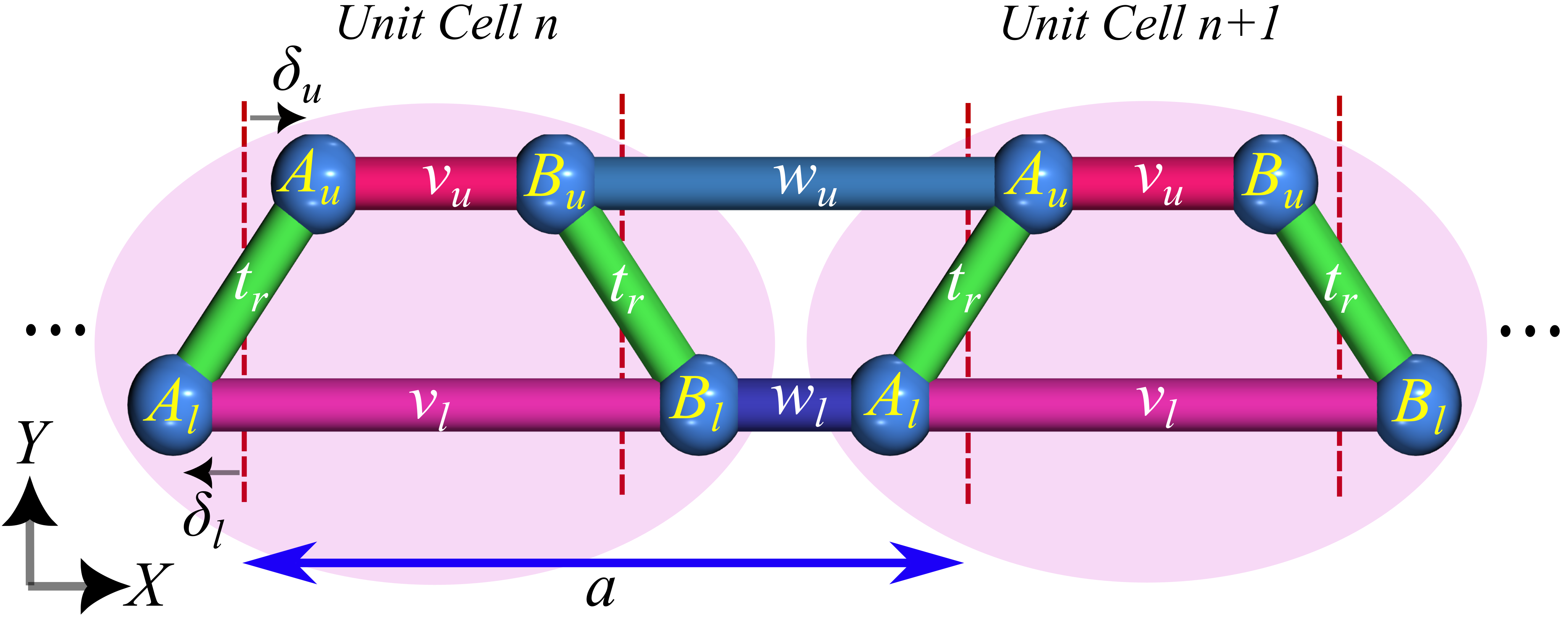}
 \caption{Schematic representation of the two-leg SSH ladder model. The upper and lower legs have unequal dimerizations, $\delta_u\neq\delta_l$. The blue double-headed arrow denotes the unit-cell length \(a\), while the dashed vertical lines mark the reference positions \(na\) and \(na+a/2\) of the undimerized lattice. The labels \(\delta_u\) and \(\delta_l\) denote the signed dimerizations of the upper and lower legs, respectively. Each unit cell contains four sites with intracell hoppings \(v_{u,l}\), intercell hoppings \(w_{u,l}\), and interleg hoppings \(t_r\).}
 \label{fig:System}
\end{figure}

\section{Model and Velocity Gauge}

We consider a two-leg ladder system composed of two coupled
SSH chains, where each lattice site carries an
onsite potential $V_{s_L}$. Each leg, labeled
by $L\equiv u,l$, consists of $N$ unit cells with two sublattice sites
$s\equiv A,B$ per unit cell. We use the ordered Bloch basis
\(\Psi=(A_u,B_u,A_l,B_l)\). Using the
tensor-product convention
$\Psi=L\otimes s$, the first factor acts in the leg space \((u,l)\),
while the second acts in the sublattice space \((A,B)\).
The position-aware field-free
Hamiltonian in this basis is
~\cite{moos_intense-laser-driven_2020}

\begin{equation}
h(k)=
\begin{pmatrix}
V_{A_u} & g_u(k) & r_A(k) & 0\\
g_u^*(k) & V_{B_u} & 0 & r_B(k)\\
r_A^*(k) & 0 & V_{A_l} & g_l(k)\\
0 & r_B^*(k) & g_l^*(k) & V_{B_l}
\end{pmatrix}.
\label{eq: Ham}
\end{equation}

The diagonal elements \(V_{s_L}\) are the onsite potentials of
sublattice \(s=A,B\) on leg \(L=u,l\). The off-diagonal elements
\(g_L(k)\) describe hopping along each leg, whereas \(r_A(k)\) and
\(r_B(k)\) describe hopping between the two legs on the \(A\) and
\(B\) sublattices, respectively. Explicitly,

\begin{equation*}
\begin{aligned}
g_L(k)&=v_L e^{ikd_{v,L}}+w_L e^{-ikd_{w,L}}, \\
r_A(k)&=t_r e^{-ik\Delta},\qquad
r_B(k)=t_r e^{ik\Delta}=r_A^*(k),
\end{aligned}
\end{equation*}

where \(v_L=-e^{-d_{v,L}}\) and \(w_L=-e^{-d_{w,L}}\) are the intracell and intercell hopping
amplitudes along leg \(L\), respectively, and \(t_r\) is the interleg
hopping amplitude. The phases in these matrix elements encode the
longitudinal displacements of the corresponding bonds. The longitudinal bond lengths along leg \(L\) are

\begin{equation}
d_{v,L}=\frac{a}{2}-2\delta_L,
\qquad
d_{w,L}=\frac{a}{2}+2\delta_L,
\qquad
d_{v,L}+d_{w,L}=a,
\label{eq:lattice-geometry}
\end{equation}

where \(a\) is the unit-cell length shown in Fig.~\ref{fig:System},
\(\delta_L\) is the signed dimerization of leg \(L\), and we set \(a=2\) for numerical calculations. The geometric and hopping parameters are chosen for illustrative purposes and are not intended to model a specific material quantitatively.
The relative longitudinal displacement of the two legs is characterized by

\begin{equation*}
\Delta=\delta_u-\delta_l,
\end{equation*}

so that the \(A\)- and \(B\)-rung displacements are \(+\Delta\) and
\(-\Delta\), respectively.

 Atomic units are used throughout ($\hbar = |e| = m_e = 1$). With this convention, $\omega_0 = 0.024$ corresponds to a photon energy of $0.653$~eV, i.e., a driving wavelength $\lambda \simeq 1.90~\mu\mathrm{m}$. Spin is not included. Since the Hamiltonian is spin-independent and the dynamics are non-interacting, including spin would only multiply the currents by a factor of two and leaves all selection rules and symmetry-based cancellations unchanged. Quarter and half filling correspond to one and two occupied bands, respectively, out of the four bands of the ladder.
Thus, $\delta_L>0$ shortens the intracell bond and lengthens the intercell bond, while $\delta_L<0$ does the opposite. We assume $|\delta_L|<a/4$, so that both bond lengths remain positive.

For the clean ladder, i.e., with all onsite potentials set to zero, let \(\widehat{\mathcal R}_x\) denote the real-space reflection
\[
x\rightarrow a/2-x
\]
about the axis \(x=a/4\). It maps the lattice sites according to
\[
(n,A_L)\leftrightarrow(-n,B_L),
\qquad L=u,l,
\]
while leaving the leg index unchanged.

In the Bloch representation, the same spatial reflection maps
\(k\rightarrow -k\). Its action on the four orbital components
\((A_u,B_u,A_l,B_l)\) is represented by
\[
R_x=I\otimes\sigma_x,
\]
which exchanges \(A\leftrightarrow B\) within each leg and therefore maps the \(A\)-rung onto the \(B\)-rung. Consequently, the field-free Bloch Hamiltonian satisfies
\[
R_x h(k)R_x^{-1}=h(-k).
\]

The clean ladder also has particle-hole symmetry,
\[
\mathcal{P}h(k)\mathcal{P}^{-1}=-h(-k),
\qquad
\mathcal{P}=P\mathcal{K}
=\sigma_z\otimes\sigma_z\mathcal{K}.
\]
Here, \(I\) is the identity matrix, \(\sigma_x\), \(\sigma_y\), and
\(\sigma_z\) are Pauli matrices, and \(\mathcal K\) denotes complex conjugation. In
addition, Hamiltonian~\eqref{eq: Ham} exhibits the combined
anti-unitary symmetry
\(\mathcal{S}=R_x\mathcal{P}
=S\mathcal{K}
=\sigma_z\otimes\sigma_y\mathcal{K}\), obtained by composing
reflection and particle-hole symmetry, such that
\(\mathcal{S}h(k)\mathcal{S}^{-1}=-h(k)\).
Both \(\mathcal P\) and \(\mathcal S\) are anti-unitary. Their unitary
parts satisfy

\begin{eqnarray*}
&P h^*(k)P^{-1}
= -h(-k), \qquad Sh^*(k)S^{-1} = -h(k).
\end{eqnarray*}

Note that the clean ladder is bipartite, with sublattices \(\{A_u,B_l\}\) and \(\{B_u,A_l\}\), since the rungs connect \(A\)--\(A\) and \(B\)--\(B\) sites. It therefore also possesses a unitary chiral symmetry$$
\Gamma=\sigma_z\otimes\sigma_z,
\qquad
\Gamma h(k,t)\Gamma^{-1}=-h(k,t),
$$which is the unitary part of \(\mathcal{P}\). Although $\Gamma$ anticommutes with $h$ just as $S$ does,
the anti-unitarity of $S$, rather than the unitarity of $\Gamma$, turns out to be essential: We return to this point in Sec.~III, once the density-matrix formalism is introduced.

For comparison, full spatial inversion reverses both the
longitudinal and transverse coordinates. In Bloch space it also maps
\(k\to-k\), while its action on the four orbital components is
represented by
\[
\Pi=\sigma_x\otimes\sigma_x,
\]
which exchanges both \(A\leftrightarrow B\) and
\(u\leftrightarrow l\). Inversion therefore requires
\(v_u=v_l\) and \(w_u=w_l\), equivalently
\(\delta_u=\delta_l\). The asymmetric ladder considered here breaks
full inversion while preserving the reflection
\(\widehat{\mathcal R}_x\).

Throughout this paper, we find that both $R_x$ and $\mathcal{S}$ symmetries play crucial roles in determining the selection rules, and therefore we restrict our analysis to these two symmetries. To examine the role of symmetry, we apply three types of onsite potentials.
 \begin{itemize}
 \item Potential I: breaks $\mathcal S$ and preserves $R_x$:\\ $V_{A_u}=V_{B_u}=-V_{A_l}=-V_{B_l}=V$.
 \item Potential II: breaks $R_x$ and preserves $\mathcal S$:\\ $V_{A_u}=V_{A_l}=-V_{B_u}=-V_{B_l}=V$,
  \item Potential III: breaks both $R_x$ and $\mathcal S$:\\ $V_{A_u}=-V_{A_l}=-V_{B_u}=-V_{B_l}=V$,
\end{itemize}
where potential I is a static transverse (inter-leg) bias between the two legs, while Potential II is a staggered, Rice–Mele-type sublattice potential~\cite{rice1982elementary}, and Potential III is a leg-asymmetric staggered potential.

\begin{figure}[t]
 \centering
\includegraphics[width=\columnwidth]{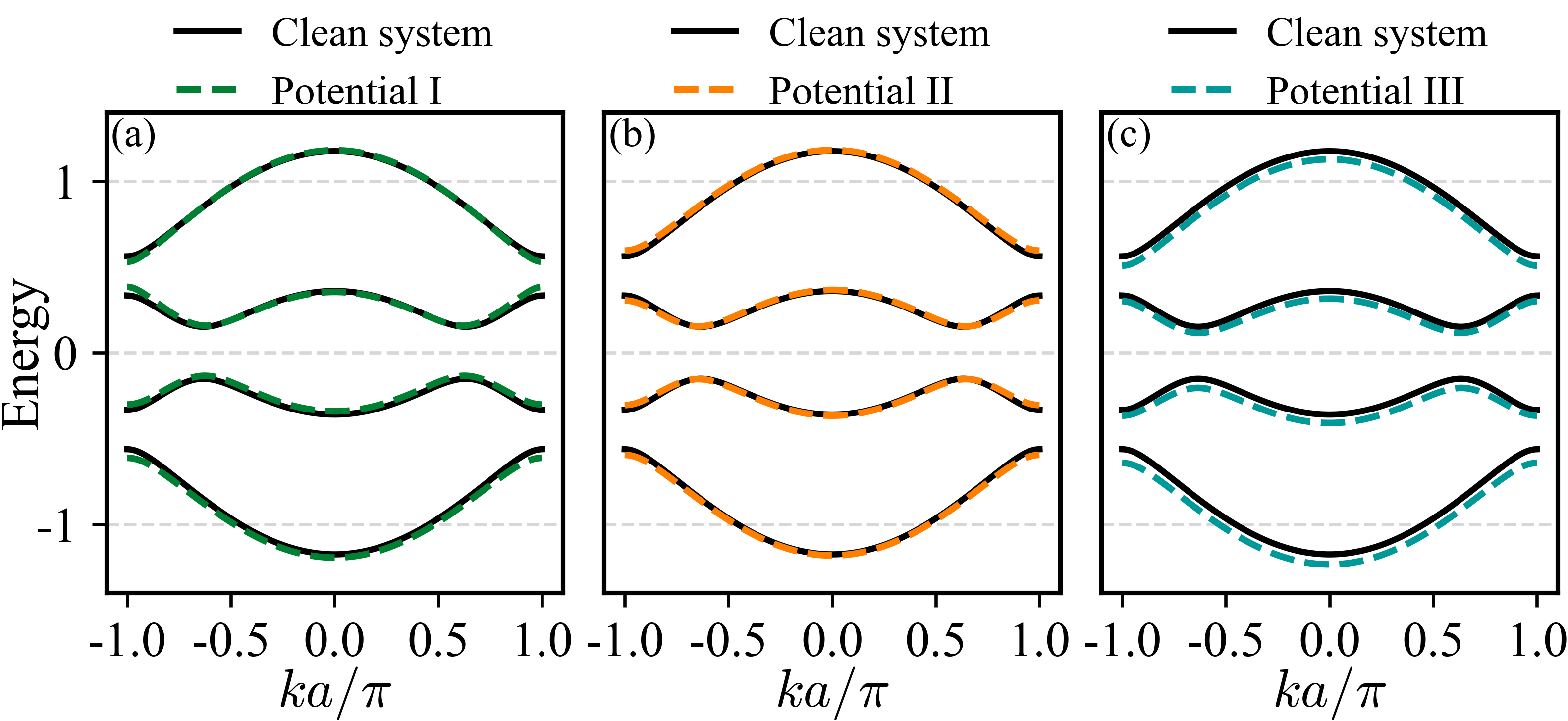}
 \caption{Band structures of the clean ladder, shown by black solid lines, compared with those obtained with (a) Potential~I  (b) Potential~II, (c) Potential~III, shown by dashed lines. Here, $\delta_u=0.05, \delta_l=-0.2$, $t_r=v_u$, and $V=0.1$.}
 \label{fig:BandS}
\end{figure}

The band structures with and without onsite potentials are shown in Fig.~\ref{fig:BandS}. The clean ladder, shown by the black solid lines in all three panels, has bands symmetric about zero energy and preserves both $R_x$ and $\mathcal S$. Potential I [panel (a)], shown by green dashed lines, breaks particle-hole symmetry and $\mathcal S$ while preserving $R_x$. Potential II [panel (b)], shown by orange dashed lines, breaks particle-hole symmetry and $R_x$ while preserving $\mathcal S$. Finally, Potential~III, shown by the cyan dashed lines in panel (c), breaks both $R_x$ and $S$.

\subsection{Velocity-gauge formulation and current operators}

The interaction between the laser field and the lattice system is described within the velocity gauge. We consider a spatially uniform laser field linearly polarized along the ladder direction \(x\), with vector potential
\begin{equation*}
\mathbf A(t)=A_x(t)\hat{\mathbf x}.
\end{equation*}
Since the system is periodic only along \(x\), the field enters the Bloch Hamiltonian through the minimal-coupling prescription
\[
k\rightarrow q(t)=k+A_x(t).
\]
We consider a 10-cycle flat-top pulse~\cite{navarrete2019crystal}, where

\begin{eqnarray}
&A_x(t)= A_0\sin(\omega_0 t)f(t),  \\
 & f(t)=
\begin{cases}
t/2T, & 0\leq t\leq 2T\ \\
1, & 2T<t\leq 8T\ \\
(10T-t)/2T, & 8T<t\leq 10T
\end{cases}.
\end{eqnarray}

Here, $f(t)$ is the dimensionless pulse envelope, rising linearly from $0$ to $1$ over the first two cycles, remaining constant at the plateau value $1$ over the following six cycles, and decreasing linearly back to $0$ over the final two cycles. Thus, $A_0$ is directly the plateau vector-potential amplitude. Here, $\omega_0$ is the driving frequency, and $T = 2\pi/\omega_0$ is the optical period. For $A_0=1.1$ and $\omega_0=0.024$, the peak electric-field amplitude on the flat-top part of the pulse is $E_0=A_0\omega_0=0.0264$ a.u., corresponding to an intensity of approximately $2.5\times10^{13}\,\mathrm{W/cm^2}$, which lies within the range used in strong-field solid-state HHG experiments.

For a hopping amplitude \(t_{ij}\) between sites \(i\) and \(j\), the effect of the driving field is incorporated through the Peierls substitution,

\begin{equation}
t_{ij}\rightarrow
t_{ij}\exp\left[-i\int_{\mathbf r_i}^{\mathbf r_j}
\mathbf A(t)\cdot d\mathbf l\right],
\label{eq:Peierls}
\end{equation}

where $\mathbf r_i$ and $\mathbf r_j$ denote the lattice-site positions. Here, $t_{ij}$ multiplies $\hat{c}_j^\dagger \hat{c}_i$, the operator that hops an electron from site $i$ to site $j$, so that the phase in Eq.~(\ref{eq:Peierls}) is attached to hopping from $i$ to $j$. This convention is used consistently in Appendix~\ref{app:currents}. Accordingly, only bonds with a finite $x$ projection acquire Peierls phases. Consequently, the slanted interleg hoppings become

\begin{equation*}
r_A(q)=t_r e^{-iq\Delta},\qquad
r_B(q)=t_r e^{iq\Delta},
\end{equation*}

with $q=k+A_x(t)$.

The longitudinal current operator is obtained from the driven Bloch Hamiltonian as

\begin{equation*}
\hat J_x(k,t)=
-\left.
\frac{\partial h(q)}{\partial q}
\right|_{q=k+A_x(t)}.
\end{equation*}

Since translational invariance is absent in the transverse direction, the transverse current operator \(\hat J_y(k,t)\) is derived from the continuity equation rather than from a momentum derivative; its explicit form is given in Appendix~\ref{app:Jy}. The explicit matrix representation of \(\hat J_x(k,t)\) is given in Appendix~\ref{app:Jx}. We set the interleg separation to unity. For identical legs, \(\delta_u=\delta_l\), with \(V_{A_u}=V_{A_l}\) and \(V_{B_u}=V_{B_l}\), leg-exchange symmetry forces \(\langle J_y(t)\rangle=0\). We therefore focus on the asymmetric case \(\delta_u\neq\delta_l\).

The density matrix evolves according to

\begin{equation*}
i\dot{\rho}(k,t)
=
\left[
h(k+A_x(t)),
\rho(k,t)
\right],
\end{equation*}

and the longitudinal and transverse currents are evaluated as

\begin{equation}
\begin{aligned}
\langle J_x(t)\rangle &= \frac{1}{N_k}\sum_k \mathrm{Tr}\!\left[\rho(k,t)\hat J_x(k,t)\right],\\
\langle J_y(t)\rangle &= \frac{1}{N_k}\sum_k \mathrm{Tr}\!\left[\rho(k,t)\hat J_y(k,t)\right],
\end{aligned}
\label{eq:ExpecJ}
\end{equation}

where \(N_k\) is the number of sampled crystal momenta.

The polarization-resolved HHG spectra are obtained from the Fourier transforms of the current components,

\begin{eqnarray*}
 P_{x(y)}(\omega) = \left| \int dt \, e^{-i \omega t}
\langle J_{x(y)}(t)\rangle \right|^2.
\end{eqnarray*}

A Hanning window is applied to \(\langle J_{x,y}(t)\rangle\) before the Fourier transform to suppress spectral leakage arising from the finite pulse.

\section{Density Matrix and Symmetry Constraints}

The density operator describes the quantum state of the system $\rho(k,t)$, which evolves from an initial equilibrium state $\rho(k,0)$ according to the von Neumann equation,

\begin{equation*}
i\frac{d\rho(k,t)}{dt}
=[h(k,t),\rho(k,t)],
\end{equation*}

where we set \(\hbar=1\), and the initial density matrix is constructed from the eigenstates of the
undriven Hamiltonian $h(k)$. Denoting its eigenvalues and
eigenvectors by $E_n(k)$ and $|n,k\rangle$, respectively, the equilibrium density matrix is given by

\begin{equation*}
\rho(k,0)=\sum_n f_n(k)|n,k\rangle\langle n,k|.
\end{equation*}

In the numerical calculations, $f_n(k)$ is a zero-temperature band occupation: at quarter filling only the lowest band is occupied, whereas at half filling the two lowest bands are occupied.

We now use the symmetries identified above to derive constraints on the time-dependent density matrix and, consequently, on the harmonic response.

Under periodic driving, the reflection symmetry $R_x$ of the static system acquires a dynamical character. For a monochromatic drive, the vector potential satisfies $A_x(t+T/2)=-A_x(t)$, and the driven Hamiltonian obeys

\begin{align}
R_x h(k,t) R_x^{-1}
= h\left(-k,t+\tfrac{T}{2}\right).
\label{eq:Rx_h}
\end{align}

Thus, Eq.~\eqref{eq:Rx_h} combines the static reflection $k\to-k$ with a half-period translation of the drive. For the finite pulse used in the numerical calculations, this relation holds within the constant-envelope part of the pulse. The ramp regions break the ideal half-period dynamical symmetry, so the corresponding selection rules are expected to hold approximately, with symmetry-forbidden harmonics strongly suppressed rather than identically absent.

For the ideal periodic response, and for an occupied Floquet subspace closed under reflection, Eq.~\eqref{eq:Rx_h} yields the density-matrix relation derived in Appendix~\ref{app:symmetry},

\begin{equation}
R_x \rho(k,t) R_x^{-1} = \rho(-k,t+T/2).
\label{eq:Rx_rho}
\end{equation}

The combined anti-unitary symmetry $\mathcal S$, when preserved, is an exact local-in-time symmetry of the driven Hamiltonian,

\begin{equation}
\mathcal S h(k,t)\mathcal S^{-1}=-h(k,t).
\label{eq:S_h}
\end{equation}

At half filling, $\mathcal S$ maps the two occupied bands onto the two unoccupied bands, and the initial density matrix therefore satisfies

\begin{equation}
\mathcal S\rho(k,0)\mathcal S^{-1}
= \mathbb I_4-\rho(k,0),
\label{eq:S_rho0}
\end{equation}

where \(\mathbb I_4\) is the four-dimensional identity. As shown in Appendix~\ref{app:symmetry}, the equation of motion preserves this relation, yielding

\begin{equation}
\mathcal S\rho(k,t)\mathcal S^{-1}
=\mathbb I_4-\rho(k,t).
\label{eq:S_rho}
\end{equation}

It is worth asking why the anti-unitary \(\mathcal{S}\), rather than the unitary chiral symmetry \(\Gamma=\sigma_z\otimes\sigma_z\) noted in Sec.~II, is required to obtain Eq.~(\ref{eq:S_rho}). At \(t=0\), \(\Gamma\) likewise maps the two occupied bands onto the two empty bands, so$$
\Gamma\rho(k,0)\Gamma^{-1}=I_4-\rho(k,0)
$$holds equally well for the clean system. This relation is not preserved by the dynamics, however, because \(\Gamma\) is unitary, \(\Gamma\rho(k,t)\Gamma^{-1}\) evolves under the generator \(-h(k,t)\), while \(I_4-\rho(k,t)\) evolves under \(+h(k,t)\), so the two coincide only at \(t=0\). It is precisely this failure to persist in time, a consequence of \(\Gamma\) being unitary rather than anti-unitary, that makes the combined symmetry \(\mathcal{S}=R_x\mathcal{P}\) essential for Eq.~(\ref{eq:S_rho}) to hold at all \(t\).

Equations~\eqref{eq:Rx_rho} and \eqref{eq:S_rho} are the central density-matrix constraints used in the selection-rule analysis below; their detailed derivations are given in Appendix~\ref{app:symmetry}. The two constraints are summarized in Table~\ref{tab:rho_symmetries}.

\begin{table}[h]
\centering
\caption{Density-matrix constraints used in the selection-rule analysis. For the ideal periodic response with a reflection-symmetric occupied set, $R_x$ gives the half-period relation independently of the number of occupied bands, whereas the local-in-time $\mathcal S$ relation requires half filling.}
\label{tab:rho_symmetries}
\renewcommand{\arraystretch}{1.3}
\begin{tabular}{@{}lccc@{}}
\hline\hline
Symmetry & Constraint on $\rho(k,t)$ & Time shift & Filling \\
\hline
$R_x$ &
$R_x\rho(k,t)R_x^{-1}=\rho(-k,t+T/2)$ &
$T/2$ & Any \\
$\mathcal S$ &
$\mathcal S\rho(k,t)\mathcal S^{-1}=\mathbb I_4-\rho(k,t)$ &
None & Half \\
\hline\hline
\end{tabular}
\end{table}

\section{Selection Rules}
\label{sec:selection}

The density-matrix constraints derived above translate directly into selection rules for the harmonic response. For the ideal periodic response, reflection symmetry determines the allowed harmonic orders of the two current components: $J_x$ is odd under a half-period shift, whereas $J_y$ is even under the same shift. At half filling, preservation of $\mathcal S$ imposes the additional constraint that the transverse current vanishes identically.

\paragraph{Reflection symmetry $R_x$:}
The current operators transform under reflection as
\begin{align*}
R_x\hat J_y(k,t)R_x^{-1}
&=+\hat J_y(-k,t+T/2),\\
R_x\hat J_x(k,t)R_x^{-1}
&=-\hat J_x(-k,t+T/2).
\end{align*}
Combining these relations with the density-matrix constraint in
Eq.~\eqref{eq:Rx_rho}, the current definition in
Eq.~\eqref{eq:ExpecJ}, cyclicity of the trace, and the
$k\to-k$ symmetry of the Brillouin-zone sum gives
\begin{eqnarray*}
\langle J_y(t+T/2)\rangle = +\langle J_y(t)\rangle, \\
\langle J_x(t+T/2)\rangle = -\langle J_x(t)\rangle.
\end{eqnarray*}
The detailed derivation is given in Appendix~\ref{app:Rx}.

These relations directly imply that only even (odd) harmonics of the driving frequency
are present in the transverse (longitudinal) spectrum:

\begin{eqnarray*}
P_y(n\omega_0) = 0 \quad \text{for odd } n, \\
P_x(n\omega_0) = 0 \quad \text{for even } n.
\end{eqnarray*}

Equivalently, the reflection selection rule can be summarized as
\begin{equation*}
R_x
\quad\Longrightarrow\quad
P_x:\ \text{odd }n,
\qquad
P_y:\ \text{even }n.
\end{equation*}
This rule constrains the allowed harmonic orders of the two current components but does not by itself require either component to be nonzero. Since no filling-specific property enters the argument, the reflection selection rule applies at any filling for which the reflection constraint holds.

\paragraph{Anti-unitary symmetry $\mathcal S$ at half filling:}

When $\mathcal S$ is preserved, at half filling the anti-unitary symmetry imposes the density-matrix constraint in Eq.~\eqref{eq:S_rho}. The transverse-current operator satisfies
\begin{equation}
\mathcal S\hat J_y(k,t)\mathcal S^{-1}=+\hat J_y(k,t),
\label{eq:S_Jy}
\end{equation}
and is traceless for every $k$. This sign is specific to \(\mathcal{S}\) in Eq.~(\ref{eq:S_Jy}): the unitary chiral operator \(\Gamma\) discussed in Secs.~II and III instead gives
\begin{equation}
\Gamma\hat{J}_y(k,t)\Gamma^{-1}
=-\hat{J}_y(k,t).
\label{eq:GammaJ}
\end{equation}
At $t=0$, where
\begin{equation}
\Gamma\rho(k,0)\Gamma^{-1}
=I_4-\rho(k,0)
\label{eq:GammaRho}
\end{equation}
holds, the minus sign from
Eq.~(\ref{eq:GammaJ}) cancels the minus sign associated with Eq.~(\ref{eq:GammaRho}),
and one recovers only the trivial identity

$$
\langle J_y(0)\rangle=\langle J_y(0)\rangle
$$
rather than a vanishing condition.
Thus, band swapping alone does not enforce the transverse-current cancellation. In addition to the density-matrix constraint, the cancellation requires the current operator to be even under the relevant symmetry, as is the case for $S$, for which
$S\hat J_yS^{-1}=+\hat J_y$, but not for $\Gamma$, for which
$\Gamma\hat J_y\Gamma^{-1}=-\hat J_y$.

As shown in Appendix~\ref{app:S}, these properties imply
\begin{equation*}
\langle J_y(t)\rangle = 0 \quad \text{for all } t,
\end{equation*}
and therefore
\begin{equation*}
P_y(n\omega_0) = 0 \quad \text{for all } n.
\end{equation*}

Thus, the anti-unitary half-filling constraint can be summarized as
\begin{equation*}
\mathcal S+\text{half filling}
\quad\Longrightarrow\quad
P_y=0.
\end{equation*}

When both $R_x$ and $\mathcal S$ are preserved at half filling, the two constraints act simultaneously: reflection fixes the allowed harmonic orders, while the half-filled $\mathcal S$ constraint suppresses the transverse channel. The combined result is
\begin{equation*}
R_x+\mathcal S+\text{half filling}
\quad\Longrightarrow\quad
P_x:\ \text{odd }n,
\qquad
P_y=0.
\end{equation*}

Eq.~(\ref{eq:S_rho}) can be restated in terms of the leg-population operator
$$
M=\sigma_z\otimes I,
$$
which is traceless and satisfies
$$
\mathcal{S}M\mathcal{S}^{-1}=+M.
$$
It implies
$$
\langle M(t)\rangle=0
$$
for all \(t\), i.e.,
$$
N_u(t)=N_l(t)
$$
throughout the pulse. The two legs therefore remain exactly equally populated, so there is no net transverse charge transfer to radiate. The formal statement
$$
\langle J_y(t)\rangle=0
$$
is equivalent to this simpler picture through the relation
$$
\hat{J}_y=-\dot{\hat{N}}_u.
$$

For the longitudinal current,
\begin{equation*}
\mathcal S\hat J_x(k,t)\mathcal S^{-1}=-\hat J_x(k,t).
\end{equation*}
As shown in Appendix~\ref{app:S}, this transformation yields only an identity for $\langle J_x(t)\rangle$ and therefore imposes no additional selection rule on the longitudinal current. If $R_x$ is also preserved, the reflection-derived odd-harmonic selection rule remains valid at any filling.

\begin{figure}[t]
 \centering
\includegraphics[width=0.5\textwidth]{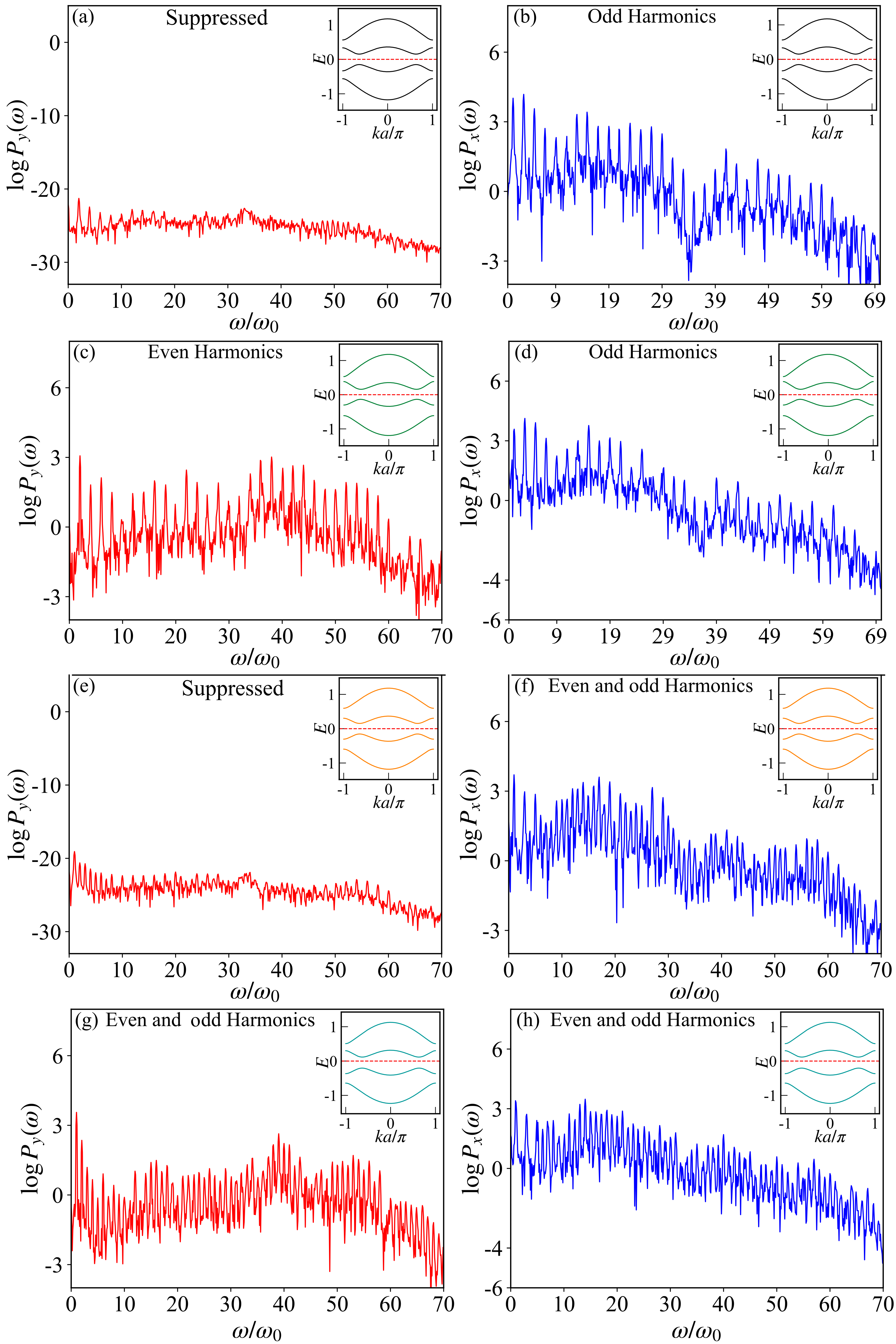}
 \caption{HHG power spectrum in the transverse (left column) and longitudinal (right column) directions at half filling. Each row represents the clean ladder ($V=0$)[(a,b)], Potential~I [(c,d)], Potential~II [(e,f)], and Potential~III [(g,h)] ($V=0.1$). Red (blue) dashed lines indicate even (odd) harmonics. Here, $\delta_u=0.05, \delta_l=-0.2$, $t_r=v_u$, $\omega_0=0.024$, and $A_0=1.1$. The inset shows the band structure, with the dashed line marking the Fermi level.}
 \label{fig:half}
\end{figure}

\begin{figure}[t]
 \centering
\includegraphics[width=0.5\textwidth]{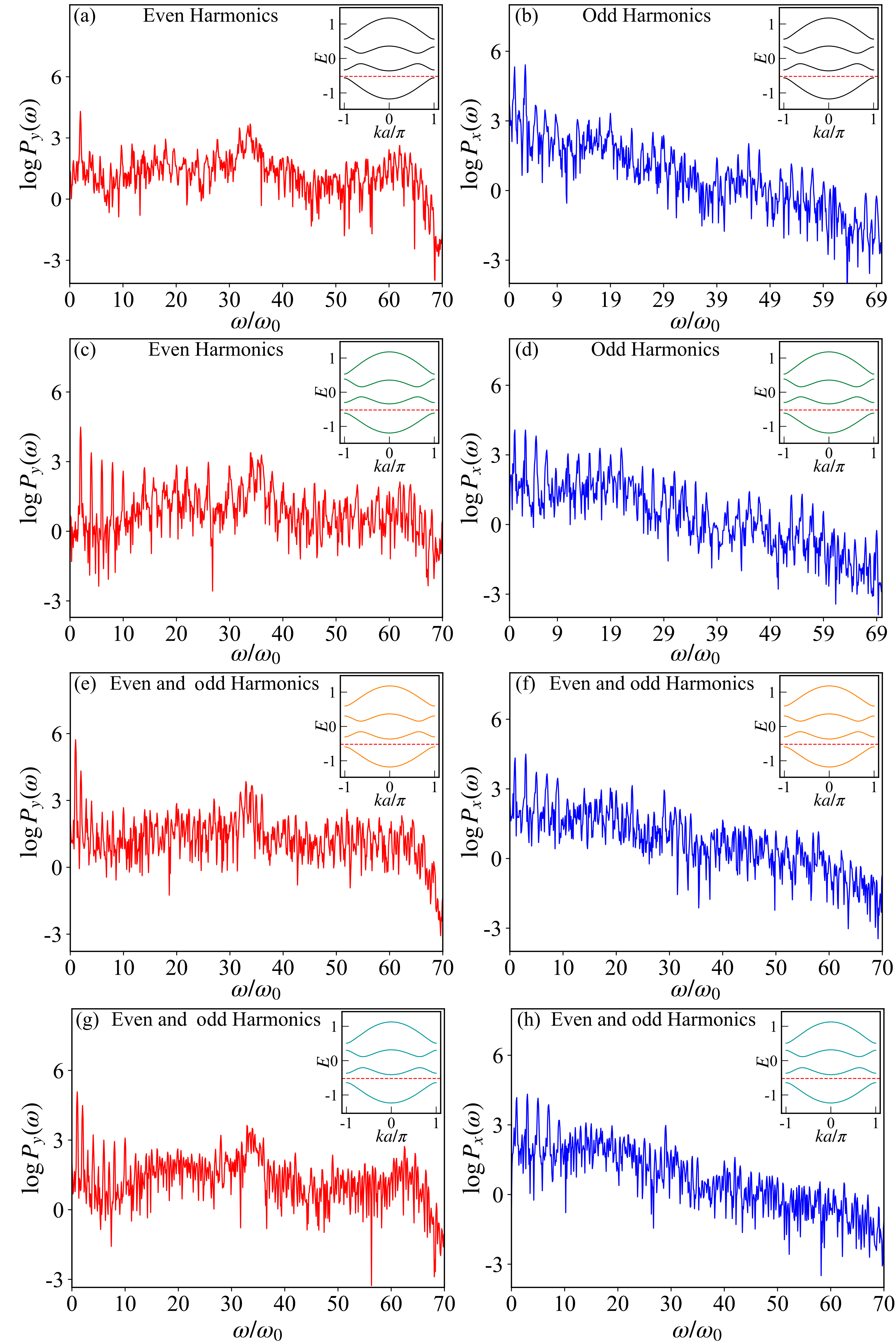}
 \caption{HHG power spectrum in the transverse (left column) and longitudinal (right column) directions at quarter filling. Each row represents the clean ladder ($V=0$)[(a,b)], Potential~I [(c,d)], Potential~II [(e,f)], and Potential~III [(g,h)] ($V=0.1$). Red (blue) dashed lines indicate even (odd) harmonics. Here, $\delta_u=0.05, \delta_l=-0.2$, $t_r=v_u$, $\omega_0=0.024$, and $A_0=1.6$. The inset shows the band structure, with the dashed line marking the Fermi level.}
 \label{fig:quarter}
\end{figure}


\section{Numerical Results}

To verify the symmetry-based selection rules derived above, we calculate the HHG spectra in the transverse ($y$) and longitudinal ($x$) directions by numerically solving the time-dependent von Neumann equation. In all calculations, \(\delta_u=0.05\), \(\delta_l=-0.2\), and \(t_r=v_u\). Labeling the four bands as $E_1(k) \leq E_2(k) \leq E_3(k) \leq E_4(k)$, the minimum direct gap $\min_k(E_2-E_1)$, relevant at quarter filling, and $\min_k(E_3-E_2)$, relevant at half filling, remain open at every $k$ for all four cases considered. $\min_k(E_2-E_1) = 0.23$, $0.31$, $0.29$, $0.28$ and $\min_k(E_3-E_2) = 0.30$, $0.29$, $0.31$, $0.32$, for the clean ladder and Potentials~I, II, and III, respectively. We consider two insulating fillings of the four-band ladder: quarter filling, where only the lowest band is initially occupied, and half filling, where the two lowest bands are initially occupied.
Because quarter and half filling involve different relevant band pairs, the
field-driven trajectories sample different interband dispersions. Since the
drive enters through $q(t)=k+A_x(t)$, the vector-potential amplitude controls
the momentum-space excursion and therefore the range of interband energy
separations explored during the pulse. We use $A_0=1.6$ at quarter filling
and $A_0=1.1$ at half filling to obtain well-developed spectra over comparable
cutoff ranges, while keeping the driving frequency fixed at $\omega_0=0.024$.
Within each filling, the same driving parameters are used for the clean ladder
and all three onsite potentials.
The Brillouin zone is sampled with \(N_k=200\) points on a mesh symmetric under \(k\to -k\), as required for the derivation in Appendix.~\ref{app:Rx} to apply to the discretized momentum sum. The von Neumann equation is propagated using the unitary Crank--Nicolson (Cayley) scheme~\cite{crank1996practical,askar1978explicit}
$$ \rho(t+\Delta t)=U\rho(t)U^\dagger$$ with $$U=\left(I+\frac{i}{2}H\Delta t\right)^{-1}
\left(I-\frac{i}{2}H\Delta t\right),
$$
which preserves Hermiticity and the trace of \(\rho\) to machine precision at every step. The time step is $\Delta t = 0.05$. The reported spectra were checked to be converged with respect to both $N_k$ and $\Delta t$.

Figure~\ref{fig:half} shows the half-filling spectra and Fig.~\ref{fig:quarter} the quarter-filling spectra for all four cases considered: the clean ladder [(a,b)], Potential~I [(c,d)], Potential~II [(e,f)], and Potential~III [(g,h)], with the transverse spectrum $P_y$ in the left column and the longitudinal spectrum $P_x$ in the right column of each row.

We first consider the clean ladder. Figure~\ref{fig:half}(a,b) shows the
spectra at half filling. The transverse signal [Fig.~\ref{fig:half}(a)]
is numerically negligible, while the longitudinal spectrum
[Fig.~\ref{fig:half}(b)] is dominated by odd-order harmonics. This agrees with the selection rules derived above: at half filling, $\mathcal{S}$ suppresses the transverse current, while $R_x$ restricts the longitudinal response to odd harmonics.

We next consider Potential I, which preserves $R_x$ but
breaks $\mathcal{S}$. Because the $\mathcal{S}$-based cancellation no longer applies, a nonzero transverse response appears at half filling
[Fig.~\ref{fig:half}(c)], in contrast to the clean ladder. The transverse spectrum is dominated by even harmonics, while
the longitudinal response remains dominated by odd harmonics [Fig.~\ref{fig:half}(d)], as expected from the preserved reflection symmetry. This shows that the suppression of the
transverse response in the clean half-filled system originates from $\mathcal{S}$, rather than from half filling alone.

We then consider Potential II, which preserves $\mathcal{S}$
but breaks $R_x$. At half filling, the transverse HHG signal remains numerically negligible [Fig.~\ref{fig:half}(e)] despite
the absence of reflection symmetry, consistent with the
$\mathcal{S}$-based suppression of the transverse current. In contrast, the longitudinal spectrum is no longer restricted
to odd harmonics and contains both even and odd orders
[Fig.~\ref{fig:half}(f)]. We note further that $\Gamma$ is not a symmetry of the Potential-II Hamiltonian.
For the chiral operator defined above, any nonzero diagonal onsite potential breaks the chiral symmetry, whereas $S$ requires only $V_{BL}=-V_{AL}$ on each leg. Potential II satisfies the latter condition with $V\neq0$, and therefore preserves $S$ while breaking $\Gamma$.
The fact that the transverse signal remains suppressed at half filling despite the complete absence of
chiral symmetry provides direct numerical confirmation that $S$, rather than chirality, controls the suppression.
Thus, Potential II provides a direct numerical separation of the two symmetries: the suppression survives when $S$ is preserved even though $\Gamma$ is broken.

Finally, we consider Potential III, which breaks both
$R_x$ and $\mathcal{S}$. At half filling, the transverse HHG signal is no
longer suppressed [Fig.~\ref{fig:half}(g)], since neither symmetry-based mechanism remains to protect it. The spectrum
contains both even- and odd-order harmonics. The longitudinal spectrum likewise exhibits both even- and odd-order harmonics [Fig.~\ref{fig:half}(h)], reflecting the breaking of
$R_x$.

At quarter filling, the half-filling constraint associated
with $\mathcal{S}$ no longer applies. As shown in Fig.~\ref{fig:quarter}(a,b), a nonzero
transverse HHG signal appears, with dominant peaks at
even harmonic orders [Fig.~\ref{fig:quarter}(a)]. The longitudinal spectrum remains dominated by odd harmonics [Fig.~\ref{fig:quarter}(b)].
Thus, changing the filling removes the suppression of the
transverse current while leaving the $R_x$ selection rule intact. For Potential I, Fig.~\ref{fig:quarter}(c,d) shows the same reflection-controlled harmonic structure: the transverse response
is nonzero and dominated by even harmonics, while the
longitudinal spectrum is dominated by odd harmonics.
These results are again consistent with the preserved $R_x$
symmetry. For Potential II, the half-filling $\mathcal{S}$ constraint no longer
applies and the transverse response becomes nonzero
[Fig.~\ref{fig:quarter}(e)]. Since $R_x$ is broken, neither current component is restricted to a single set of harmonic orders,
and both even and odd harmonics appear in the transverse and longitudinal spectra [Fig.~\ref{fig:quarter}(f)]. Finally, for Potential III, the same generic structure persists [Fig.~\ref{fig:quarter}(g,h)], with both current components containing
even- and odd-order harmonics. This is the expected
behavior when both symmetry mechanisms are absent:
neither the reflection selection rule nor the anti-unitary
suppression constrains the spectra, leaving no filling- or
symmetry-based restriction on the harmonic content.

Table~\ref{tab:selection_main} summarizes the harmonic content obtained in all eight numerical cases. The results agree with the selection rules derived above: when $R_x$ is preserved, the longitudinal response is dominated by odd harmonics and a nonzero transverse response by even harmonics; when $R_x$ is broken, both even and odd harmonic orders can occur; and when $\mathcal S$ is preserved at half filling, the transverse signal is suppressed.

\begin{table}[h]
\centering
\small
\setlength{\tabcolsep}{4pt}
\caption{Observed harmonic content of the numerically calculated spectra for the asymmetric ladder, $\delta_u\neq\delta_l$. The symmetries shown in parentheses indicate which of $R_x$ and $\mathcal S$ are preserved in each case. ``Odd $n$'' and ``Even $n$'' denote spectra dominated by the corresponding harmonic orders for the finite pulse, ``Even/odd $n$'' indicates visible contributions from both even and odd orders, and ``Suppressed'' denotes a numerically negligible harmonic signal.}
\label{tab:selection_main}
\renewcommand{\arraystretch}{1.4}
\begin{tabular}{@{}llcc@{}}
\hline\hline
System (symmetry) & Filling & $P_x(n\omega_0)$ & $P_y(n\omega_0)$ \\
\hline
Clean $(R_x,\mathcal S)$
    & Quarter & Odd $n$      & Even $n$ \\
Clean $(R_x,\mathcal S)$
    & Half    & Odd $n$      & Suppressed \\
Pot.\ I $(R_x)$
    & Quarter & Odd $n$      & Even $n$ \\
Pot.\ I $(R_x)$
    & Half    & Odd $n$      & Even $n$ \\
Pot.\ II $(\mathcal S)$
    & Quarter & Even/odd $n$ & Even/odd $n$ \\
Pot.\ II $(\mathcal S)$
    & Half    & Even/odd $n$ & Suppressed \\
Pot.\ III (none)
    & Quarter   & Even/odd $n$ & Even/odd $n$ \\  
Pot.\ III (none)
    & Half    & Even/odd $n$ & Even/odd $n$ \\
\hline\hline
\end{tabular}
\end{table}

\FloatBarrier

\section{Summary}\label{s8}

We have investigated HHG in the longitudinal ($x$) and transverse ($y$) directions of a laser-driven two-leg SSH ladder with asymmetric dimerization under periodic boundary conditions. Since edge-state contributions are excluded by construction, the resulting selection rules originate entirely from bulk symmetries. On an open chain, zero-energy edge modes are not necessarily paired by \(\mathcal{S}\) with an unoccupied partner, so the pairing argument underlying Eq.~(\ref{eq:S_rho0}) may fail and the transverse suppression could be lifted. This provides a testable prediction for finite ladders, which we leave for future work.

Two distinct symmetry mechanisms govern the harmonic response. Reflection symmetry $R_x$, together with the half-period sign reversal of the drive, restricts the longitudinal response to odd harmonic orders and the transverse response to even harmonic orders. This selection rule constrains the allowed harmonic orders but does not by itself require a nonzero transverse current. The asymmetric dimerization considered here removes the leg-exchange symmetry that, for identical legs, would force the transverse current to vanish.

At half filling, preservation of the combined anti-unitary symmetry $\mathcal S=R_x\mathcal P$ imposes an additional constraint, $\langle J_y(t)\rangle=0$, and therefore suppresses transverse HHG altogether. Away from half filling, this cancellation no longer applies, allowing a transverse response while the reflection selection rule remains whenever $R_x$ is preserved.

Numerical calculations at quarter and half filling, together with onsite potentials that selectively break $R_x$ or $\mathcal S$, confirm these predictions. These results confirm the two distinct symmetry mechanisms outlined in the Introduction.

The symmetry \(\mathcal{S}\) relies on strictly bipartite hopping and is broken by any next-nearest-neighbor hopping along a leg or by a diagonal rung term. Thus, the on/off switch described here is a statement about an idealized model rather than a generic material; possible realizations include cold-atom, photonic, or acoustic ladder lattices, where such couplings can be engineered to vanish. We note also that a relaxation channel respecting \(\mathcal{S}\) would leave the suppression argument intact, whereas a symmetry-breaking bath would not.

\section*{Acknowledgments}
We thank M. V. Hosseini for useful comments. M.J. gratefully acknowledges funding by the Deutsche Forschungsgemeinschaft (DFG, German Research Foundation) through IRTG 2676/1 ‘Imaging of Quantum Systems’, project no. 437567992. F.N. and D.B. gratefully acknowledge funding from the German Research Foundation (DFG), SFB 1477 ‘Light-Matter Interactions at Interfaces’, project number 441234705.

\clearpage

\appendix


\section{Bloch convention and current operators}
\label{app:currents}

In this appendix, we construct the driven Hamiltonian and both current
operators using the position-aware Bloch convention adopted in the main
text. We first compare it with the conventional Bloch convention and then
derive the longitudinal and transverse current operators.

\subsection{Bloch conventions in the field-free and driven problems}
\label{app:bloch-conventions}

In the ordered basis

\begin{equation*}
\Psi_k=(A_u,B_u,A_l,B_l)^T,
\end{equation*}

the field-free Hamiltonian in the conventional Bloch basis is

\begin{equation*}
\bar h(k)=
\begin{pmatrix}
V_{A_u} & \bar g_u(k) & t_r & 0\\
\bar g_u^*(k) & V_{B_u} & 0 & t_r \\
t_r  & 0 & V_{A_l} & \bar g_l(k)\\
0 & t_r & \bar g_l^*(k) & V_{B_l}
\end{pmatrix},
\end{equation*}

 with

\begin{equation}
\bar g_L(k)=v_L+w_Le^{-ika},
\qquad L=u,l,
\label{app:eq:conventional-leg}
\end{equation}

 where \(a\) is the translation length of one unit cell, as indicated
in Fig.~\ref{fig:System}. In this convention, the intracell orbital
positions are omitted from the Bloch phases. 

For the geometry used in the main text, the longitudinal orbital
positions inside the unit cell are
\begin{equation*}
\tau_{A_L}^{x}=\delta_L,
\qquad
\tau_{B_L}^{x}=\frac{a}{2}-\delta_L,
\qquad L=u,l.
\end{equation*}
These positions reproduce the bond lengths in Eq.~\eqref{eq:lattice-geometry}.
We then define

 \begin{equation*}
U(k)=
\operatorname{diag}
\left(
e^{ik\tau_{A_u}^{x}},
e^{ik\tau_{B_u}^{x}},
e^{ik\tau_{A_l}^{x}},
e^{ik\tau_{B_l}^{x}}
\right).
\end{equation*}

The position-aware Hamiltonian is

\begin{equation*}
h(k)=U^\dagger(k)\bar h(k)U(k).
\end{equation*}

Therefore the two field-free Hamiltonians have identical eigenvalues, provided the same numerical values of \(v_L\), \(w_L\), and \(t_r\) are used:

\begin{equation*}
E_n^{\rm pos}(k)=E_n^{\rm conv}(k).
\end{equation*}

With these orbital positions and the bond lengths in
Eq.~\eqref{eq:lattice-geometry}, the directed displacements entering
the Bloch phases are \(+d_{v,L}\) for the intracell hopping and
\(-d_{w,L}\) for the intercell hopping. The position-aware matrix
element is then

\begin{equation}
g_L(k)=v_Le^{ikd_{v,L}}+w_Le^{-ikd_{w,L}}=e^{ikd_{v,L}}\bar g_L(k).
\label{app:eq:position-aware-leg-comparison}
\end{equation}

 Equation~\eqref{app:eq:position-aware-leg-comparison} makes the
 field-free equivalence explicit.

In the position-aware convention, a longitudinal vector potential is
 introduced directly by

\begin{equation*}
q(t)=k+A_x(t),
\end{equation*}

 so that

 \begin{equation*}
g_L(q)=v_Le^{iqd_{v,L}}+w_Le^{-iqd_{w,L}}.
\end{equation*}

By contrast, the naive replacement in
 Eq.~\eqref{app:eq:conventional-leg},

\begin{equation*}
\bar g_L(k+A_x)
=
v_L+w_Le^{-i(k+A_x)a},
\end{equation*}

 couples only the intercell term to the field.

The conventional Bloch basis remains valid, but the correctly Peierls-coupled matrix element is

\begin{equation*}
\bar g_L(k,A_x)
=
v_Le^{iA_xd_{v,L}}
+
w_Le^{-ika}e^{-iA_xd_{w,L}},
\end{equation*}

which satisfies
\begin{equation*}
g_L(k+A_x)=e^{ikd_{v,L}}\bar g_L(k,A_x).
\end{equation*}

In addition, the rung elements are affected in the same way. In the conventional Bloch basis, the field-free rung hopping is simply $t_r$, independent of $k$, since Eq.~(A1) does not carry any phase for the rung term. The physical rungs, however, carry longitudinal displacements $\pm\Delta$. The Peierls phase must therefore be attached to the rung hopping directly in real space, before Fourier transforming, giving the driven rung amplitude $t_r e^{\mp iA_x\Delta}$; only after this substitution is the Hamiltonian transformed to $k$-space. This phase cannot instead be introduced afterward by any substitution in $k$ applied to the field-free $\bar{h}(k)$, since the rung hopping $t_r$ in $\bar{h}(k)$ carries no $k$-dependence into which the field could be substituted.

In the position-aware convention, this gives the simple prescription
\begin{equation*}
h(k,t)=h\bigl(k+A_x(t)\bigr).
\end{equation*}

We follow the Bloch convention used in Appendix A of
Ref.~\cite{moos_intense-laser-driven_2020}, in which the orbital
positions are included in the Bloch phases; here we refer to it as the
position-aware Bloch convention.

\subsection{Real-space and position-aware Bloch Hamiltonians}
\label{app:real-space-Hamiltonian}

Using the geometry and hopping amplitudes defined in Eq.~\eqref{eq:lattice-geometry}, the field-free real-space Hamiltonian is

\begin{align*}
\hat H_0
={}&\sum_{n,L}
\Bigl[
V_{A_L}\hat n_{nA_L}
+V_{B_L}\hat n_{nB_L}
+\Bigl(
 v_L\hat c^\dagger_{nA_L}\hat c_{nB_L}
\\
&\hspace{2.8cm}
+w_L\hat c^\dagger_{n+1,A_L}\hat c_{nB_L}
+\mathrm{H.c.}
\Bigr)
\Bigr]
\\
&+t_r\sum_n
\Bigl(
\hat c^\dagger_{nA_u}\hat c_{nA_l}
+\hat c^\dagger_{nB_u}\hat c_{nB_l}
+\mathrm{H.c.}
\Bigr),
\end{align*}

 where \(L=u,l\), $n$ labels the unit cell, and $\hat n_{n\alpha}=\hat c^\dagger_{n\alpha}\hat c_{n\alpha}$. This equation fixes unambiguously which hopping is
intracell, which hopping is intercell, and which sites are connected by
the rungs. The conventional and position-aware Bloch Hamiltonians below
are two representations of this same real-space operator.

Using the position-aware Bloch transformation

\begin{equation*}
\hat c_{n\alpha}
=
\frac{1}{\sqrt{N}}
\sum_k
e^{ik(na+\tau_{\alpha}^{x})}
\hat c_{k\alpha},
\end{equation*}

the leg matrix elements are

\begin{equation*}
g_L(k)
=
v_Le^{ikd_{v,L}}
+
w_Le^{-ikd_{w,L}},
\end{equation*}

with the bond lengths defined in Eq.~\eqref{eq:lattice-geometry}.

The position-aware rung matrix elements are

\begin{equation*}
r_A(k)=t_re^{-ik\Delta},
\qquad
r_B(k)=t_re^{ik\Delta}=r_A^*(k),
\end{equation*}

where \(\Delta=\delta_u-\delta_l\) is defined in the main text and
\(t_r\) is taken to be real.

The complete field-free Hamiltonian is

\begin{equation*}
h(k)=
\begin{pmatrix}
V_{A_u} & g_u(k) & r_A(k) & 0\\
g_u^*(k) & V_{B_u} & 0 & r_B(k)\\
r_A^*(k) & 0 & V_{A_l} & g_l(k)\\
0 & r_B^*(k) & g_l^*(k) & V_{B_l}
\end{pmatrix}.
\end{equation*}

For a field polarized along \(x\), the driven Hamiltonian used in the time evolution is

\begin{equation*}
h(k,t)=h\bigl(q(t)\bigr),
\qquad q(t)=k+A_x(t).
\end{equation*}

\subsection{Longitudinal current}
\label{app:Jx}

In the position-aware convention, every longitudinal bond displacement
is contained in the \(q\)-dependence of the Hamiltonian. The complete
longitudinal electronic-current operator is

\begin{equation*}
J_x(k,t)
=
-\left.
\frac{\partial h(q)}{\partial q}
\right|_{q=k+A_x(t)}.
\end{equation*}

The overall sign corresponds to electron charge \(-1\).

For the leg matrix elements,

 \begin{equation*}
j_L^x(q)
=
-i d_{v,L}v_Le^{iqd_{v,L}}
+i d_{w,L}w_Le^{-iqd_{w,L}}.
\end{equation*}

For the interleg bonds,

\begin{equation*}
j_A^x(q)=i\Delta r_A(q),
\qquad
j_B^x(q)=-i\Delta r_B(q).
\end{equation*}

The complete current matrix is

\begin{equation*}
J_x(k,t)
=
\begin{pmatrix}
0 & j_u^x(q) & j_A^x(q) & 0\\
[j_u^x(q)]^* & 0 & 0 & j_B^x(q)\\
[j_A^x(q)]^* & 0 & 0 & j_l^x(q)\\
0 & [j_B^x(q)]^* & [j_l^x(q)]^* & 0
\end{pmatrix}.
\end{equation*}

Here \(q=k+A_x(t)\).

This matrix contains the intracell, intercell, and longitudinal projections of the slanted interleg currents.

\subsection{Transverse current from the continuity equation}
\label{app:Jy}

 Because translational invariance is absent in the transverse
direction, there is no transverse crystal momentum \(k_y\). We derive
the transverse current directly from the continuity equation.

 Define the directed hopping operator from the lower leg to the upper
leg in position space as

\begin{align*}
\hat h_{\rm rung}(t)
=
t_r\sum_n
\Bigl[
&e^{-iA_x(t)\Delta}
\hat c^\dagger_{nA_u}\hat c_{nA_l}
\\
&+
e^{iA_x(t)\Delta}
\hat c^\dagger_{nB_u}\hat c_{nB_l}
\Bigr].
\end{align*}

The Hermitian rung Hamiltonian is

\begin{equation*}
\hat H_{\rm rung}(t)
=
\hat h_{\rm rung}(t)+\hat h_{\rm rung}^\dagger(t).
\end{equation*}

 The total upper-leg occupation is

\begin{equation*}
\hat N_u
=
\sum_n
\left(\hat n_{nA_u}+\hat n_{nB_u}\right).
\end{equation*}

 Only the rung Hamiltonian changes \(\hat N_u\).

 Using

\begin{equation*}
\dot{\hat N}_u
=
-i[\hat N_u,\hat H_{\rm rung}]
\end{equation*}

 and

\begin{equation*}
[\hat N_u,\hat h_{\rm rung}]=\hat h_{\rm rung},
\qquad
[\hat N_u,\hat h_{\rm rung}^\dagger]=-\hat h_{\rm rung}^\dagger.
\end{equation*}

 one obtains

\begin{equation*}
\dot{\hat N}_u
=
-i\hat h_{\rm rung}
+i\hat h_{\rm rung}^\dagger.
\end{equation*}

For electron charge \(-1\), and using the interleg separation as the
unit of transverse length as in the main text, the transverse charge
current in the positive \(y\) direction is

\begin{equation*}
\hat J_y
=
-\dot{\hat N}_u
=
i\,\hat h_{\rm rung}
-
i\,\hat h_{\rm rung}^\dagger.
\end{equation*}

Applying the position-aware Bloch transformation gives

 \begin{align*}
\hat h_{\rm rung}(t)
&=\sum_k\Bigl[
r_A(q)\hat c^\dagger_{kA_u}\hat c_{kA_l}
+r_B(q)\hat c^\dagger_{kB_u}\hat c_{kB_l}
\Bigr],\\
q&=k+A_x(t).
\end{align*}

 Therefore,

\begin{equation*}
J_y(k,t)
=
\begin{pmatrix}
0 & 0 & i r_A(q) & 0\\
0 & 0 & 0 & i r_B(q)\\
-i r_A^*(q) & 0 & 0 & 0\\
0 & -i r_B^*(q) & 0 & 0
\end{pmatrix}.
\end{equation*}

Here \(q=k+A_x(t)\).

 The explicit \(k\) dependence follows from the physical longitudinal
offsets of the slanted rungs and must be retained together with the
position-aware density matrix.

In the identical-leg limit \(\delta_u=\delta_l\),
Eq.~\eqref{eq:lattice-geometry} gives identical bond lengths and
hoppings on the two legs, and therefore
\begin{equation*}
g_u(q)=g_l(q).
\end{equation*}
At the same time, \(\Delta=0\), so that
\begin{equation*}
r_A(q)=r_B(q)=t_r.
\end{equation*}
If, in addition, \(V_{A_u}=V_{A_l}\) and \(V_{B_u}=V_{B_l}\), the
Hamiltonian is invariant under interchange of the upper and lower legs.
Such an interchange reverses the transverse direction and therefore
changes the sign of \(\hat J_y\). For an initial state respecting the
same symmetry, the time evolution preserves it, and consequently
\(\langle J_y(t)\rangle=0\).

\section{Symmetry constraints on the density matrix}\label{app:symmetry}

In this appendix we derive the density-matrix constraints used in the main text. We separate the reflection-based dynamical symmetry, which relies on the half-period relation of a periodic driver, from the local-in-time anti-unitary symmetry $\mathcal{S}$, which gives the half filling cancellation.

\subsection*{Reflection Symmetry $R_x$}

For the field-free initial state, reflection symmetry implies
\begin{equation*}
R_x\rho(k,0)R_x^{-1}=\rho(-k,0).
\end{equation*}

For the ideal periodic response considered below, we assume that the occupied
Floquet subspace is continuously connected to the occupied field-free
subspace. Since the latter is closed under reflection, the corresponding
occupied Floquet subspace is chosen to be closed under the dynamical
reflection symmetry derived below.

For a strictly periodic monochromatic drive, the vector potential satisfies

\begin{equation*}
A_x(t+T/2)=-A_x(t).
\end{equation*}

Together with reflection symmetry, this gives

\begin{equation}
R_x h(k,t)R_x^{-1}=h(-k,t+T/2).
\label{eq:app:Rx_h}
\end{equation}

Let

\begin{align*}
\lvert\psi_{m k}(t)\rangle
&=e^{-i\varepsilon_{m k}t}\lvert u_{m k}(t)\rangle,\\
\lvert u_{m k}(t+T)\rangle
&=\lvert u_{m k}(t)\rangle.
\end{align*}

Here \(m\) labels a Floquet quasienergy branch. The periodic part obeys

\begin{equation*}
{
\left[h(k,t)-i\partial_t\right]
\lvert u_{m k}(t)\rangle
=
\varepsilon_{m k}
\lvert u_{m k}(t)\rangle .
}
\end{equation*}

Applying \(R_x\) and using Eq.~\eqref{eq:app:Rx_h},

\begin{equation*}
{
\left[h(-k,t+T/2)-i\partial_t\right]
R_x\lvert u_{m k}(t)\rangle
=
\varepsilon_{m k}
R_x\lvert u_{m k}(t)\rangle .
}
\end{equation*}

Hence, for a nondegenerate quasienergy branch and consistent branch labels at \(k\) and \(-k\),

\begin{equation}
{
R_x\lvert u_{m k}(t)\rangle
=
e^{i\chi_{m k}}
\lvert u_{m,-k}(t+T/2)\rangle .
}
\label{eq:app:Rx_floquet}
\end{equation}

Here, $\chi_{mk}$ is an arbitrary relative phase between the two
symmetry-related Floquet states and therefore drops out of the corresponding
projector.
This Floquet--Bloch symmetry relation follows the constructions of Refs.~\cite{Moiseyev2015SelectionSolids,Neufeld2019FloquetGroup}. At a quasienergy degeneracy, the corresponding statement applies to a symmetry-adapted projector onto the degenerate subspace. For each occupied Floquet--Bloch orbital, define

\begin{equation*}
{
\rho_{m k}(t)
=
\lvert\psi_{m k}(t)\rangle
\langle\psi_{m k}(t)\rvert
=
\lvert u_{m k}(t)\rangle
\langle u_{m k}(t)\rvert .
}
\end{equation*}

Using Eq.~\eqref{eq:app:Rx_floquet},

\begin{align*}
R_x\rho_{m k}(t)R_x^{-1}
&=R_x\lvert u_{m k}(t)\rangle
\langle u_{m k}(t)\rvert R_x^{-1}\\
&=e^{i\chi_{m k}}
\lvert u_{m,-k}(t+T/2)\rangle\\
&\quad\times
\langle u_{m,-k}(t+T/2)\rvert e^{-i\chi_{m k}}\\
&=\rho_{m,-k}(t+T/2).
\end{align*}

The one-particle density matrix is

\begin{equation*}
{
\rho(k,t)
=
\sum_{m\in\mathrm{occ}}
\rho_{m k}(t).
}
\end{equation*}

For an occupied set mapped onto itself by reflection,

\begin{align*}
{
R_x\rho(k,t)R_x^{-1}
}
&{
=
\sum_{m\in\mathrm{occ}}
R_x\rho_{m k}(t)R_x^{-1}
}
\\
&{
=
\sum_{m\in\mathrm{occ}}
\rho_{m,-k}(t+T/2).
}
\end{align*}

Thus, summing over the occupied Floquet branches proves the
density-matrix constraint stated in Eq.~\eqref{eq:Rx_rho}.

Equation~\eqref{eq:Rx_rho} is exact for the ideal periodic Floquet response provided the occupied set is closed under reflection; it does not depend on the number of occupied branches. For the finite pulse, the ramp can generate superpositions of Floquet branches, so the relation is approximate during the plateau and the forbidden harmonic orders are strongly suppressed rather than identically zero.

We next consider the combined local-in-time anti-unitary symmetry \(\mathcal{S}=R_x\mathcal{P}\).

\subsection*{Anti-Unitary Symmetry $\mathcal{S}$}

The combined anti-unitary symmetry $\mathcal{S} = \sigma_z\otimes\sigma_y\mathcal{K}$ acts locally in time. It satisfies

\begin{equation}
\mathcal{S}h(k,t)\mathcal{S}^{-1}=-h(k,t).
\label{eq:app:S_h}
\end{equation}

Because \(\mathcal{S}\) is anti-unitary, applying it to the von Neumann equation changes the sign of \(i\). Defining

\begin{equation*}
\bar\rho_S(k,t)=\mathcal{S}\rho(k,t)\mathcal{S}^{-1},
\end{equation*}

we obtain

\begin{align*}
-i\partial_t\bar\rho_S(k,t)
&=
\left[
\mathcal{S}h(k,t)\mathcal{S}^{-1},
\bar\rho_S(k,t)
\right]
\\
&=
\left[
-h(k,t),
\bar\rho_S(k,t)
\right].
\end{align*}

Equivalently,

\begin{equation}
i\partial_t\bar\rho_S(k,t)
=
\left[
h(k,t),
\bar\rho_S(k,t)
\right].
\label{eq:app:S_rho_eom}
\end{equation}

Hence \(\mathcal{S}\rho(k,t)\mathcal{S}^{-1}\) evolves with the same Liouvillian as \(\rho(k,t)\).

The remaining input is the initial state. At half filling, the two lowest bands are occupied and the two upper bands are empty. Since Eq.~\eqref{eq:app:S_h} maps an eigenstate of energy \(\varepsilon_n(k)\) to one of energy \(-\varepsilon_n(k)\), \(\mathcal{S}\) maps the initially filled bands onto the initially empty bands. This pairing requires the spectrum to be gapped at \(E=0\) for all \(k\), so that the occupied and unoccupied subspaces are unambiguously separated and mapped onto one another by \(\mathcal{S}\). Therefore

\begin{equation}
\mathcal{S}\rho(k,0)\mathcal{S}^{-1}
=
\mathbb{I}_4-\rho(k,0).
\label{eq:app:S_rho0}
\end{equation}

The matrix \(\mathbb{I}_4-\rho(k,t)\) obeys the same equation of motion as \(\rho(k,t)\), because

\begin{equation*}
i\partial_t\left[\mathbb{I}_4-\rho(k,t)\right]
=
\left[
h(k,t),\mathbb{I}_4-\rho(k,t)
\right].
\end{equation*}

Combining this with Eqs.~\eqref{eq:app:S_rho_eom} and \eqref{eq:app:S_rho0}, uniqueness of the solution proves the density-matrix constraint stated in Eq.~\eqref{eq:S_rho}.

This relation is exact for any pulse shape, as long as the driven Hamiltonian preserves \(\mathcal{S}\) and the initial state satisfies Eq.~\eqref{eq:app:S_rho0}. Away from half filling, the initially filled and empty sectors are not completely paired by \(\mathcal{S}\), so Eq.~\eqref{eq:app:S_rho0} fails and the constraint \eqref{eq:S_rho} no longer follows.

\section{Selection Rules from Symmetry: Detailed Derivations}
\label{app:selection}

In this appendix, we derive the harmonic selection rules for the transverse ($J_y$) and longitudinal ($J_x$) currents. The current expectation values are defined in Eq.~\eqref{eq:ExpecJ}. The key symmetry actors are the reflection symmetry $R_x$, which acts at any filling, and the combined anti-unitary symmetry $\mathcal{S}=R_x\mathcal{P}$, whose density-matrix constraint is operative only at half filling.

For a strictly periodic drive, each current component can be expanded in a Fourier series,

\begin{align*}
J_{x,y}(t)
&=\sum_{n=-\infty}^{\infty} J_n\, e^{-in\omega_0 t},\\
J_n
&=\frac{1}{T}\int_0^T dt\, J_{x,y}(t)\, e^{in\omega_0 t}.
\end{align*}

Here $\omega_0 = 2\pi/T$ is the driving frequency, and the selection rules correspond to identifying which Fourier coefficients $J_n$ must vanish. For the finite pulse used in the numerical calculations, the corresponding rules appear as suppression of the symmetry-forbidden harmonic components, up to spectral broadening from the envelope.

\subsection{Dynamical reflection symmetry $R_x$ at any filling:
 odd harmonics in $J_x$, even harmonics in $J_y$}
\label{app:Rx}

We now use the reflection constraint in Eq.~\eqref{eq:Rx_rho} to determine the allowed harmonic orders of the two current components.

\paragraph{Transverse current $J_y$.}
The operator $\hat{J}_y$ transforms as

\begin{equation}
R_x\hat{J}_y(k,t)R_x^{-1}
=
+\hat{J}_y(-k,t+T/2)
\label{app:eq:Rx_Jy}
\end{equation}

Using Eqs.~\eqref{eq:Rx_rho} and \eqref{app:eq:Rx_Jy} together with
the cyclic property of the trace and the $k\to -k$ symmetry of the
Brillouin-zone sum,

\begin{align*}
\langle J_y(t+T/2)\rangle
&= \frac{1}{N_k}\sum_k
 \mathrm{Tr}\bigl[\rho(k,t+T/2)\,\hat{J}_y(k,t+T/2)\bigr]
\\
&= \frac{1}{N_k}\sum_k
 \mathrm{Tr}\bigl[R_x\rho(-k,t)\hat{J}_y(-k,t)R_x^{-1}\bigr]
\\
&= \frac{1}{N_k}\sum_k
 \mathrm{Tr}\bigl[\rho(-k,t)\,\hat{J}_y(-k,t)\bigr]
\\
&= +\langle J_y(t)\rangle.
\end{align*}

The current is therefore \emph{periodic} with half-period $T/2$.
Substituting into the Fourier series,

\begin{eqnarray*}
J_y(t+T/2) = \sum_n J_n\,e^{-in\omega_0(t+T/2)} \\
 =
 \sum_n J_n\,(-1)^n\,e^{-in\omega_0 t}
 \stackrel{!}{=} \sum_n J_n\,e^{-in\omega_0 t}.
\end{eqnarray*}

Matching coefficients gives $J_n(-1)^n = J_n$, which requires $(-1)^n = 1$, i.e., $n$ must be even. Therefore

\begin{equation*}
J_n = 0 \quad \text{for odd } n,
\end{equation*}

and the transverse current contains only even harmonics whenever it is nonzero and \(R_x\) is preserved.

\paragraph{Longitudinal current $J_x$.}
The operator $\hat{J}_x$ transforms with the opposite sign,

\begin{equation*}
R_x\hat{J}_x(k,t)R_x^{-1} = -\hat{J}_x(-k,\,t+T/2)
\end{equation*}

The identical calculation now gives an overall minus sign,

\begin{align*}
\langle J_x(t+T/2)\rangle
&= -\frac{1}{N_k}\sum_k
 \mathrm{Tr}\bigl[R_x\rho(-k,t)\hat{J}_x(-k,t)R_x^{-1}\bigr]
\\
&= -\frac{1}{N_k}\sum_k
 \mathrm{Tr}\bigl[\rho(-k,t)\,\hat{J}_x(-k,t)\bigr]
\\
&= -\langle J_x(t)\rangle.
\end{align*}

The longitudinal current is \emph{anti-periodic} with half-period $T/2$.
Substituting into the Fourier series gives
$J_n(-1)^n = -J_n$, requiring $(-1)^n = -1$, i.e.\ $n$ odd. Therefore

\begin{equation*}
J_n = 0 \quad \text{for even } n,
\end{equation*}

and, when $R_x$ is preserved, the longitudinal current contains only odd harmonics at any
filling. This is the standard result for systems with inversion-like
symmetry combined with a half-period shift.

\subsection{Anti-unitary symmetry $\mathcal{S}=R_x\mathcal{P}$ at half
 filling: complete suppression of $J_y$}
\label{app:S}

At half filling, we use the density-matrix constraint in Eq.~\eqref{eq:S_rho}. For the zero-temperature band projector used here, this relation holds because $\mathcal S$ maps the occupied two-band subspace onto its unoccupied complement. Half filling alone would not be sufficient for an arbitrary density matrix. Away from half filling, the initially occupied and empty subspaces are not completely paired by $\mathcal{S}$ and the constraint does not follow.

The transverse current operator transforms under $\mathcal{S}$ as

\begin{equation}
\mathcal{S}\,\hat{J}_y(k,t)\,\mathcal{S}^{-1}
=
+\hat{J}_y(k,t).
\label{app:eq:S_Jy}
\end{equation}

We now evaluate $\langle J_y(t)\rangle$ using the transformed density
matrix. Writing the anti-unitary operator as $\mathcal S=S\mathcal K$, with $S$ unitary, one has
$\mathcal S A\mathcal S^{-1}=SA^*S^\dagger$ and therefore
$\mathrm{Tr}[\mathcal S A\mathcal S^{-1}]=\mathrm{Tr}[A]^*$.
Using Eqs.~\eqref{eq:S_rho} and \eqref{app:eq:S_Jy},

\begin{align*}
\langle J_y(t)\rangle
&= \frac{1}{N_k}\sum_k
\mathrm{Tr}\bigl[\rho(k,t)\hat{J}_y(k,t)\bigr]
\\
&=
\left\{
\frac{1}{N_k}\sum_k
\mathrm{Tr}\bigl[
\mathcal{S}\rho(k,t)\mathcal{S}^{-1}
\mathcal{S}\hat{J}_y(k,t)\mathcal{S}^{-1}
\bigr]
\right\}^{*}
\\
&=
\left\{
\frac{1}{N_k}\sum_k
 \mathrm{Tr}\bigl[
 \bigl(\mathbb{I}_4 - \rho(k,t)\bigr)\hat{J}_y(k,t)
 \bigr]
\right\}^{*}.
\end{align*}

Because both $\mathbb I_4-\rho(k,t)$ and $\hat J_y(k,t)$ are Hermitian,
the trace of their product is real:
$\mathrm{Tr}[AB]^*=\mathrm{Tr}[(AB)^\dagger]
=\mathrm{Tr}[BA]=\mathrm{Tr}[AB]$.
Hence,

\begin{align*}
\langle J_y(t)\rangle
&= \frac{1}{N_k}\sum_k
 \Bigl(
 \mathrm{Tr}\bigl[\hat{J}_y(k,t)\bigr]
 - \mathrm{Tr}\bigl[\rho(k,t)\hat{J}_y(k,t)\bigr]
 \Bigr)
\\
&= \frac{1}{N_k}\sum_k
\mathrm{Tr}\bigl[\hat{J}_y(k,t)\bigr]
- \langle J_y(t)\rangle.
\end{align*}

From the explicit matrix representation in Appendix~\ref{app:Jy},
\begin{equation*}
\mathrm{Tr}\bigl[\hat{J}_y(k,t)\bigr]=0
\qquad \text{for every }k.
\end{equation*}

Therefore,

\begin{equation*}
\langle J_y(t)\rangle = -\langle J_y(t)\rangle
\quad\Longrightarrow\quad
\langle J_y(t)\rangle = 0 \quad \text{for all } t,
\end{equation*}

and consequently

\begin{equation*}
P_y(\omega) = 0 \quad \text{for all harmonic orders.}
\end{equation*}

Thus, when $\mathcal S$ is preserved at half filling, the transverse HHG signal is completely suppressed.

\paragraph{Why $\mathcal{S}$ does not constrain $J_x$.}
For the longitudinal current, $\mathcal{S}$ transforms with the opposite sign,

\begin{equation*}
\mathcal{S}\,\hat{J}_x(k,t)\,\mathcal{S}^{-1} = -\hat{J}_x(k,t).
\end{equation*}

Repeating the same calculation gives

\begin{equation*}
\langle J_x(t)\rangle
=
-\frac{1}{N_k}\sum_k
\mathrm{Tr}\bigl[\hat{J}_x(k,t)\bigr]
+\langle J_x(t)\rangle.
\end{equation*}

Since
\begin{equation*}
\mathrm{Tr}\bigl[\hat{J}_x(k,t)\bigr]=0
\qquad \text{for every }k,
\end{equation*}
this reduces to the identity
\begin{equation*}
\langle J_x(t)\rangle=\langle J_x(t)\rangle.
\end{equation*}
Thus, $\mathcal{S}$ imposes no constraint on the longitudinal current. If $R_x$ is also preserved, the reflection-derived odd-harmonic selection rule remains valid at any filling; when $R_x$ is broken, no restriction to even or odd harmonic orders in $J_x$ follows from these symmetries.

\clearpage

\bibliography{References}

@article{Ghimire2011Observation,
  title = {Observation of high-order harmonic generation in a bulk crystal},
  author = {Ghimire, Shambhu and DiChiara, Anthony D. and Sistrunk, Emily and Agostini, Pierre and DiMauro, Louis F. and Reis, David A.},
  journal = {Nature Physics},
  volume = {7},
  number = {2},
  pages = {138--141},
  year = {2011},
  publisher = {Nature Publishing Group},
  doi = {10.1038/nphys1847}
}

@article{Vampa2015Theoretical,
  title = {Theoretical analysis of high-harmonic generation in solids},
  author = {Vampa, G. and McDonald, C. R. and Orlando, G. and Klug, D. D. and Corkum, P. B. and Brabec, T.},
  journal = {Physical Review Letters},
  volume = {113},
  number = {7},
  pages = {073901},
  year = {2014},
  publisher = {American Physical Society},
  doi = {10.1103/PhysRevLett.113.073901}
}

@article{Vampa2015Linking,
  title = {Linking high harmonics from gases and solids},
  author = {Vampa, G. and Hammond, T. J. and Thir{\'e}, N. and Schmidt, B. E. and L{\'e}gar{\'e}, F. and McDonald, C. R. and Brabec, T. and Corkum, P. B.},
  journal = {Nature},
  volume = {522},
  number = {7557},
  pages = {462--464},
  year = {2015},
  publisher = {Nature Publishing Group},
  doi = {10.1038/nature14517}
}

@article{Hohenleutner2015RealTime,
  title = {Real-time observation of interfering crystal electrons in high-harmonic generation},
  author = {Hohenleutner, M. and Langer, F. and Schubert, O. and Knorr, M. and Huttner, U. and Koch, S. W. and Kira, M. and Huber, R.},
  journal = {Nature},
  volume = {523},
  number = {7562},
  pages = {572--575},
  year = {2015},
  publisher = {Nature Publishing Group},
  doi = {10.1038/nature14652}
}

@article{Luu2015Extreme,
  title = {Extreme ultraviolet high-harmonic spectroscopy of solids},
  author = {Luu, T. T. and Garg, M. and Kruchinin, S. Yu. and Moulet, A. and Hassan, M. Th. and Goulielmakis, E.},
  journal = {Nature},
  volume = {521},
  number = {7553},
  pages = {498--502},
  year = {2015},
  publisher = {Nature Publishing Group},
  doi = {10.1038/nature14456}
}

@article{Schubert2014Subcycle,
  title = {Sub-cycle control of terahertz high-harmonic generation by dynamical {Bloch} oscillations},
  author = {Schubert, O. and Hohenleutner, M. and Langer, F. and Urbanek, B. and Lange, C. and Huttner, U. and Golde, D. and Meier, T. and Kira, M. and Koch, S. W. and Huber, R.},
  journal = {Nature Photonics},
  volume = {8},
  number = {2},
  pages = {119--123},
  year = {2014},
  publisher = {Nature Publishing Group},
  doi = {10.1038/nphoton.2013.349}
}

@article{Luu2018Measurement,
  title = {Measurement of the {Berry} curvature of solids using high-harmonic spectroscopy},
  author = {Luu, T. T. and W{\"o}rner, H. J.},
  journal = {Nature Communications},
  volume = {9},
  number = {1},
  pages = {916},
  year = {2018},
  publisher = {Nature Publishing Group},
  doi = {10.1038/s41467-018-03397-4}
}

@article{Neufeld2019Symmetries,
  title = {Selection rules for breaking selection rules},
  author = {Even Tzur, Matan and Neufeld, Ofer and Fleischer, Avner and Cohen, Oren},
  journal = {New Journal of Physics},
  volume = {23},
  number = {10},
  pages = {103039},
  year = {2021},
  doi = {10.1088/1367-2630/ac27e2}
}

@article{Bauer2018High,
  title = {High-Harmonic Generation in Solids with and without {Topological Edge States}},
  author = {Bauer, Dieter and Hansen, Kenneth K.},
  journal = {Physical Review Letters},
  volume = {120},
  number = {17},
  pages = {177401},
  year = {2018},
  publisher = {American Physical Society},
  doi = {10.1103/PhysRevLett.120.177401}
}

@article{Ciappina2025SolidState,
  title = {Solid-state high-order harmonic generation: emerging frontiers in ultrafast and quantum light science},
  author = {Ciappina, Marcelo F.},
  journal = {Advances in Physics},
  volume = {74},
  number = {1-4},
  pages = {1--9},
  year = {2025},
  publisher = {Taylor \& Francis},
  doi = {10.1080/00018732.2025.2598130}
}

@book{Ciappina2024HighOrder,
  title = {{High-Order Harmonic Generation in Solids}},
  editor = {Ciappina, Marcelo and Tzallas, Paraskevas},
  publisher = {World Scientific},
  address = {Singapore},
  year = {2024},
  isbn = {9789811279553},
  doi = {10.1142/13506},
  note = {Foreword by Shambhu Ghimire and David Reis}
}

@article{Kozak2025Noncollinear,
  title = {Noncollinear phase-matching of high harmonic generation in solids},
  author = {Peterka, Pavel and Troj{\'a}nek, Franti{\v{s}}ek and Mal{\'y}, Petr and Koz{\'a}k, Martin},
  journal = {APL Photonics},
  volume = {9},
  number = {11},
  pages = {116110},
  year = {2024},
  publisher = {AIP Publishing},
  doi = {10.1063/5.0219911}
}

@article{Kraus2024Enhancing,
  title = {Enhancing the efficiency of high-order harmonics with two-color non-collinear wave mixing in silica},
  author = {Roscam Abbing, Sylvianne D. C. and Kuzkova, Nataliia and van der Linden, Roy and Campi, Filippo and de Keijzer, Brian and Morice, Corentin and Zhang, Zhuang-Yan and van der Geest, Maarten L. S. and Kraus, Peter M.},
  journal = {Nature Communications},
  volume = {15},
  pages = {8335},
  year = {2024},
  publisher = {Springer Nature},
  doi = {10.1038/s41467-024-52774-9}
}

@article{Kuzkova2025Boosting,
  title = {Boosting solid-state high-harmonic generation efficiency by non-collinear {XUV} wave mixing with two-color drivers},
  author = {Kuzkova, Nataliia and Roscam Abbing, Sylvianne D. C. and van der Linden, Roy and Campi, Filippo and de Keijzer, Brian and Morice, Corentin and Zhang, Zhuang-Yan and van der Geest, Maarten L. S. and Kraus, Peter M.},
  journal = {EPJ Web of Conferences},
  volume = {335},
  pages = {10003},
  year = {2025},
  doi = {10.1051/epjconf/202533510003}
}

@article{Garg2016Multi,
  title = {Multi-petahertz electronic metrology},
  author = {Garg, M. and Zhan, M. and Luu, T. T. and Lakhotia, H. and Klostermann, T. and Guggenmos, A. and Goulielmakis, E.},
  journal = {Nature},
  volume = {538},
  number = {7625},
  pages = {359--363},
  year = {2016},
  publisher = {Nature Publishing Group},
  doi = {10.1038/nature19821}
}

@article{jurss2019high,
  title = {High-harmonic generation in {Su-Schrieffer-Heeger} chains},
  author = {J{\"u}r{\ss}, Hannah and Bauer, Dieter},
  journal = {Physical Review B},
  volume = {99},
  number = {19},
  pages = {195428},
  year = {2019},
  publisher = {American Physical Society},
  doi = {10.1103/PhysRevB.99.195428}
}

@article{Ma2022PRB106125117,
  title = {Role of symmetry breaking in high-order harmonic generation from {Su-Schrieffer-Heeger} systems},
  author = {Ma, Chao and Bian, Xue-Bin and Du, Tao-Yuan},
  journal = {Physical Review B},
  volume = {106},
  number = {12},
  pages = {125117},
  year = {2022},
  doi = {10.1103/PhysRevB.106.125117}
}

@article{zhang2024high,
  title = {High-order harmonic generation from {Su-Schrieffer-Heeger} chains with edge states: Three-step processes and their interference},
  author = {Yu, Chuan},
  journal = {Physical Review A},
  volume = {110},
  number = {3},
  pages = {033113},
  year = {2024},
  publisher = {American Physical Society},
  doi = {10.1103/PhysRevA.110.033113}
}

@article{bera2023topological,
  title = {Topological phase detection through high-harmonic spectroscopy in extended {Su-Schrieffer-Heeger} chains},
  author = {Bera, Mohit Lal and Almeida, Jessica O. de and Dziurawiec, Marlena and P{\l}odzie{\'n}, Marcin and Ma{\'s}ka, Maciej M. and Lewenstein, Maciej and Grass, Tobias and Bhattacharya, Utso},
  journal = {Physical Review B},
  volume = {108},
  number = {21},
  pages = {214104},
  year = {2023},
  publisher = {American Physical Society},
  doi = {10.1103/PhysRevB.108.214104}
}

@article{pooyan2024harmonic,
  title = {Harmonic generation with topological edge states and electron-electron interaction},
  author = {Pooyan, S. and Bauer, D.},
  journal = {Phys. Rev. B},
  volume = {111},
  issue = {1},
  pages = {014308},
  numpages = {7},
  year = {2025},
  month = {Jan},
  publisher = {American Physical Society},
  doi = {10.1103/PhysRevB.111.014308},
  url = {https://link.aps.org/doi/10.1103/PhysRevB.111.014308}
}

@article{marques2016multihole,
  title = {Multihole edge states in {Su-Schrieffer-Heeger} chains with interactions},
  author = {Marques, A. M. and Dias, R. G.},
  journal = {Physical Review B},
  volume = {95},
  number = {11},
  pages = {115443},
  year = {2017},
  doi = {10.1103/PhysRevB.95.115443}
}

@article{caceres2025polaronic,
  title = {Polaronic Effect in High-Harmonic Generation},
  author = {C{\'a}ceres-Aravena, Gabriel and Bauer, Dieter},
    journal = {arXiv preprint},
  year = {2025},
  eprint = {2508.14633},
  archivePrefix = {arXiv},
  primaryClass = {cond-mat.mes-hall}
}

@article{cui2024multiplateau,
  title = {Multiplateau structure caused by spectral interference in high-order harmonic generation from disordered {Su-Schrieffer-Heeger} chains},
  author = {Chen, Jia-Xiang and Bian, Xue-Bin},
  journal = {Physical Review A},
  volume = {109},
  number = {3},
  pages = {033104},
  year = {2024},
  doi = {10.1103/PhysRevA.109.033104}
}

@article{jangjan2020floquet,
  title = {{Floquet} engineering of topological metal states and hybridization of edge states with bulk states in dimerized two-leg ladders},
  author = {Jangjan, Milad and Hosseini, Mir Vahid},
  journal = {Scientific Reports},
  volume = {10},
  number = {1},
  pages = {14256},
  year = {2020},
  publisher = {Nature Portfolio},
  doi = {10.1038/s41598-020-71196-3}
}

@article{su1979solitons,
  title = {Solitons in {Polyacetylene}},
  author = {Su, W. P. and Schrieffer, J. R. and Heeger, A. J.},
  journal = {Physical Review Letters},
  volume = {42},
  number = {25},
  pages = {1698--1701},
  year = {1979},
  publisher = {American Physical Society},
  doi = {10.1103/PhysRevLett.42.1698}
}

@article{Hansen2022Doping,
  author = {Hansen, Thomas and Madsen, Lars Bojer},
  title = {Doping effects in high-harmonic generation from correlated systems},
  journal = {Phys. Rev. B},
  volume = {106},
  pages = {235142},
  year = {2022},
  doi = {10.1103/PhysRevB.106.235142}
}

@article{Liu2016SelectionRules,
  author = {Liu, Xi and Zhu, Xiaosong and Li, Liang and Li, Yang and Zhang, Qingbin and Lan, Pengfei and Lu, Peixiang},
  title = {Selection rules of high-order harmonic generation: Symmetries of molecules and laser fields},
  journal = {Phys. Rev. A},
  volume = {94},
  pages = {033410},
  year = {2016},
  doi = {10.1103/PhysRevA.94.033410}
}

@article{Moiseyev2015SelectionSolids,
  author = {Moiseyev, Nimrod},
  title = {Selection rules for harmonic generation in solids},
  journal = {Phys. Rev. A},
  volume = {91},
  pages = {053811},
  year = {2015},
  doi = {10.1103/PhysRevA.91.053811}
}

@article{navarrete2019crystal,
  title={Crystal-momentum-resolved contributions to high-order harmonic generation in solids},
  author={Navarrete, Francisco and Ciappina, Marcelo F and Thumm, Uwe},
  journal = {Physical Review A},
  volume = {100},
  number = {3},
  pages = {033405},
  year = {2019},
  publisher = {American Physical Society},
  doi = {10.1103/PhysRevA.100.033405}
}

@article{Alon1998SelectionRules,
  title = {Selection rules for the high harmonic generation spectra},
  author = {Alon, Ofir E. and Averbukh, Vitali and Moiseyev, Nimrod},
  journal = {Physical Review Letters},
  volume = {80},
  pages = {3743--3746},
  year = {1998},
  doi = {10.1103/PhysRevLett.80.3743}
}

@article{Neufeld2019FloquetGroup,
  title = {Floquet group theory and its application to selection rules in harmonic generation},
  author = {Neufeld, Ofer and Podolsky, Daniel and Cohen, Oren},
  journal = {Nature Communications},
  volume = {10},
  pages = {405},
  year = {2019},
  doi = {10.1038/s41467-018-07935-y}
}

@article{drueke2019robustness,
  title = {Robustness of topologically sensitive harmonic generation in laser-driven linear chains},
  author = {Dr{\"u}eke, Helena and Bauer, Dieter},
  journal = {Physical Review A},
  volume = {99},
  pages = {053402},
  year = {2019},
  doi = {10.1103/PhysRevA.99.053402}
}

@article{moos_intense-laser-driven_2020,
  title = {Intense-laser-driven electron dynamics and high-order harmonic generation in solids including topological effects},
  author = {Moos, Daniel and J{\"u}r{\ss}, Hannah and Bauer, Dieter},
  journal = {Physical Review A},
  volume = {102},
  number = {5},
  pages = {053112},
  year = {2020},
  publisher = {American Physical Society},
  doi = {10.1103/PhysRevA.102.053112}
}

@article{ceccherini2001dynamical,
  title = {Dynamical symmetries and harmonic generation},
  author = {Ceccherini, F. and Bauer, D. and Cornolti, F.},
  journal = {Journal of Physics B: Atomic, Molecular and Optical Physics},
  volume = {34},
  number = {24},
  pages = {5017--5029},
  year = {2001},
  doi = {10.1088/0953-4075/34/24/305}
}

@article{rice1982elementary,
  title = {Elementary excitations of a linearly conjugated diatomic polymer},
  author = {Rice, Michael J. and Mele, Eugene J.},
  journal = {Physical Review Letters},
  volume = {49},
  number = {19},
  pages = {1455--1459},
  year = {1982},
  publisher = {American Physical Society},
  doi = {10.1103/PhysRevLett.49.1455}
}

@article{crank1996practical,
  title={A practical method for numerical evaluation of solutions of partial differential equations of the heat-conduction type},
  author={Crank, John and Nicolson, Phyllis},
  journal={Advances in Computational Mathematics},
  volume={6},
  number={1},
  pages={207--226},
  year={1996},
  publisher={Baltzer Science Publishers, Baarn/Kluwer Academic Publishers Dordrecht},
  doi={10.1007/BF02127704}
}

@article{askar1978explicit,
  title={Explicit integration method for the time-dependent {Schr{\"o}dinger} equation for collision problems},
  author={Askar, Attila and Cakmak, Ahmet S},
  journal={The Journal of Chemical Physics},
  volume={68},
  number={6},
  pages={2794--2798},
  year={1978},
  publisher={American Institute of Physics},
  doi={10.1063/1.436072}
}

@misc{Jangjan2026NESS,
  title = {Even-harmonic generation through nonequilibrium steady-state symmetry breaking},
  author = {Jangjan, Milad and Kashef, Arman and Pooyan, Siamak and Scheel, Stefan and Bauer, Dieter},
  year = {2026},
  eprint = {2607.27847},
  archivePrefix = {arXiv},
  primaryClass = {quant-ph},
  doi = {10.48550/arXiv.2607.27847}
}

@article{Lorenzo2026HighHarmonic,
  title = {High harmonic spectroscopy from lower-order to higher-order topological insulators},
  author = {Lorenzo, Bryan and Batista, Carlos and Jangjan, Milad and Kim, Dasol and Menotti, Jean and Liu, Feng and Gao, Wenlong and Ghimire, Shambhu and Granados, Camilo and Chac{\'o}n, Alexis},
  journal = {Physical Review B},
  volume = {114},
  number = {8},
  pages = {084307},
  year = {2026},
  doi = {10.1103/7zjc-4b4r}
}

\end{document}